# Design and Empirical Evaluation of a Network-Centric, On-Premises Architecture for Earth Observation Data Access

***Authors***
*João Pinelo[1*], João Gonçalves[1], Denis Willett[2], Amit Ruhela[3], Derek Steinmoeller[4], Uriel Mendoza[5], Pelumi S. Alao[6], Ronald Soares Lopes[7], Rogerio Atem de Carvalho[7], Pedro Mattos[8]*

[*] *Corresponding author:* joao.pinelo@aircentre.org

***Affiliations***
1. AIR Centre, Azores, Portugal
2. North Carolina Institute for Climate Studies (NCICS/NOAA), USA
3. Texas Advanced Computing Center (TACC), University of Texas at Austin, USA
4. University of Waterloo, Waterloo, Ontario, Canada
5. Laboratorio Nacional de Observación de la Tierra (LANOT), Universidad Nacional Autónoma de México (UNAM), Mexico
6. National Space Research and Development Agency (NASRDA), Nigeria
7. Instituto Federal Fluminense (IFF), Brazil
8. Laboratório de Métodos Computacionais em Engenharia (LAMCE), Universidade Federal do Rio de Janeiro, Brazil

**Abstract**

Earth observation (EO) programmes generate data at volumes that exceed the transfer and storage capacity of most institutional networks. Public cloud platforms address this for well-resourced organisations, but institutions across the Atlantic basin face constraints in connectivity, sovereignty and funding that make on-premises infrastructure the only viable path. Cloud-native data formats enable efficient partial reads, yet their performance depends on the bandwidth of the underlying network fabric, a dependency rarely measured in isolation. This paper presents a replicable, network-centric architecture for on-premises EO data access, evaluated at its first operational deployment: the AIR Data Centre, founding node of the Atlantic Cloud. The system comprises a MinIO object storage cluster on a 100 GbE fabric, a PostGIS metadata catalogue and an OGC API-EDR access layer. We characterise the fabric under sustained parallel load, evaluate object storage throughput for EO-representative workloads, and compare measured performance against throttled baselines on identical hardware, isolating network bandwidth as the sole variable. Multi-site replication benchmarks with partner institutions characterise the federation primitive the model depends on. Network bandwidth is the dominant constraint on storage throughput for bulk EO data access up to a threshold; beyond it, endpoint memory topology rather than capacity governs how much bandwidth a system can use. For this hardware class that threshold lies above 10 Gbps per server. Below it, network capacity alone sets what the facility can deliver; above it, the return on further network investment depends on endpoint memory provisioning, which can be deferred and bought later.

## 1. Introduction

Earth observation data volumes are growing faster than most institutions can absorb. The Copernicus programme alone generates approximately 20 TB of new data per day, and downstream products multiply that figure. For the growing number of institutions that need to work with this data — to monitor coastal zones, detect internal ocean waves, measure land deformation — the question is not whether EO data exists, but whether the infrastructure to use it is within reach.

The dominant response has been migration to public cloud. NASA's Earthdata Cloud now hosts over 100 PB on AWS; the Copernicus Data Space Ecosystem provides cloud-based access to the full Sentinel archive. NOAA's Open Data Dissemination Program (NODD) [1] provides 90 PB of data across AWS, GCP, and Azure. These platforms co-locate storage and compute, and for institutions with reliable high-bandwidth connectivity to major cloud regions, they work well. But the Atlantic region — from the Azores to West Africa, from Brazil to the Caribbean — is not well served by this model. Connectivity is constrained: the Azores connect to mainland Europe by a single submarine cable. Funding is project-based, not commercial, making sustained cloud expenditure difficult to plan. Data sovereignty is a concern for national agencies. And some capabilities — such as direct satellite downlink — require local infrastructure by definition. For these institutions, the infrastructure they can build and operate locally determines what science and services are possible.

This is not a problem unique to the Azores. It is structural across the Atlantic. Research institutions in Brazil, Portugal, West Africa, and small island states face variants of the same constraint: the need for on-premises EO data capacity that is affordable, operationally independent, and capable of supporting cloud-native access patterns. The Atlantic Cloud — a transatlantic network of on-premises data facilities initiated by the AIR Centre and its partners — was created to address this need. Each node is independently operational, built around a common architecture of S3-compatible object storage, spatial metadata catalogues, and standard OGC APIs. Nodes share data through MinIO's multi-site replication and expose services through a common API gateway. The model is designed to be replicable: an institution that builds a node gains local EO data capacity immediately, and joins a distributed infrastructure that amplifies the value of each participant.

This paper reports on the architecture and empirical performance of the first operational Atlantic Cloud node: the AIR Data Centre on Terceira island, Azores. The centre was designed and built to serve the AIR Centre's EO programmes — including an Internal Waves Service processing Sentinel-1 synthetic aperture radar (SAR) data and a deployed Direct Receiving Station for satellite downlink — under the connectivity and funding constraints described above. The most consequential design decision was the network. Commercial providers proposed 1 Gbps internal connectivity. We invested instead in a 100 GbE fabric — four ports per node, bonded — because both EO workloads and machine-learning workloads on EO data are fundamentally data-movement workloads. Scientific work is iterative by nature:

hypotheses are tested by repeated queries against the data. Researchers also work visually with EO imagery: inspecting scene after scene, building the visual baseline that lets phenomena and anomalies be recognised at all. A single 512 MiB Sentinel-2 granule takes an estimated 4.3 seconds to transfer at 1 Gbps line rate and an estimated 43 milliseconds at 100 Gbps; at the bandwidth proposed by the commercial providers, that visual workflow is inefficient, while at 100 GbE it is fluid. Cloud-native EO formats such as Zarr and COG depend on rapid iterative reads of small chunks, and ML training and inference pipelines stream those same data products at sustained rates. Both workload classes were anticipated at the 2021 procurement decision; both run on the fabric routinely today, and the fabric capacity is what makes the GPU tier deliver useful throughput rather than be I/O-starved on the same data. That difference changes what access patterns are feasible.

Despite its impact, network bandwidth is rarely treated as an independent variable in EO infrastructure evaluations. Storage benchmarks report throughput that implicitly assumes a non-limiting network. System papers often describe architectures without characterising the fabric. The result is that the most consequential hardware decision — how much to invest in network — is made without empirical guidance. For institutions considering their own deployments, whether as standalone facilities or as Atlantic Cloud nodes, that guidance is what this paper provides.

We evaluate the AIR Data Centre across two dimensions: network fabric characterisation under sustained parallel load, and object storage throughput for EO-representative data products at four object sizes. Critically, we measure storage performance at three network speeds — 1 Gbps, 10 Gbps, and 100 Gbps — on identical hardware, isolating the network as the sole variable. The 1 Gbps baseline corresponds to the commercial proposals we rejected; 10 Gbps represents the upper end of conventional institutional networking. The comparison gives prospective node operators the numbers they need to size their own investments.

**Contributions.** The specific contributions of this paper are:

1. A replicable, network-centric architecture for on-premises EO data access, designed for independent operation and federation through S3-compatible multi-site replication. We report on its deployment as the first operational node of the Atlantic Cloud.
2. Empirical characterisation of a 100 GbE storage fabric under sustained parallel load, including throughput, latency under congestion, and failover behaviour under progressive link failures.
3. Bounding the rate at which EO object storage ceases to be network-bound. We measure throughput for four EO-representative object sizes (4 MiB to 2 GiB) on the deployed 100 Gbps fabric, and for three sizes (4 MiB, 64 MiB, 512 MiB) at throttled baselines of 1 Gbps and 10 Gbps on identical hardware. Throughput tracks the per-server link rate up to approximately 10 Gbps; above that the relationship departs from linearity, and endpoint memory topology rather than memory capacity governs how much of the fabric is reachable. The throttled baselines are a counterfactual measurement rather than a degradation experiment: they reproduce, on the deployed hardware, the procurement option the facility rejected.
4. Systematic per-node benchmarking as a fault-detection method. The per-node, per-link protocol built to characterise the fabric surfaced four latent faults in the deployed cluster — a BIOS setting, a kernel TCP option, a congestion-control divergence, and contention from a co-resident production workload — none of which was visible to production monitoring, because load balancing distributes the deficit of an underperforming node across the cluster. A fabric-blind evaluation would have missed all four.
5. Characterisation of MinIO multi-site replication between the AIR Data Centre and several partner institutions across the Atlantic basin, establishing the federation primitive on which the Atlantic Cloud model depends. A first round of benchmarks across 63 sites — seven institutional partners and 56 public-cloud regions used as a comparator — is reported. Across that set, the class of network the replication path traverses, rather than round-trip time, is what sets the achievable rate.

## 2. Related Work

### 2.1 Cloud-Native Earth Observation Data Formats and Platforms

The shift toward cloud-native EO data access has been driven by formats designed for partial, parallel reads over HTTP. Cloud-Optimised GeoTIFF (COG)[2] reorganises raster data into internally tiled, pyramidal structures that support HTTP range requests, enabling clients to retrieve spatial and spectral subsets without downloading full files. Zarr [3] provides chunked, compressed, N-dimensional array storage with external metadata, and is increasingly adopted for multi-temporal and multi-dimensional EO datasets; ESA's Earth Observation Processing Framework (EOPF) [4] is transitioning Sentinel Level-1 and Level-2 products to Zarr. The SpatioTemporal Asset Catalog (STAC) [5] specification provides a complementary discovery layer, enabling machine-readable cataloguing of COG and Zarr assets. NASA's Cloud-Optimized Format Study [6] evaluated the access performance of COG, Zarr, NetCDF-4, and HDF5 from AWS S3, confirming that chunked, cloud-optimised formats substantially reduce access latency compared to archival formats. Large-scale EO platforms have adopted these formats within public cloud environments. NASA's Earthdata Cloud programme is migrating over 100 PB of EOSDIS data to AWS, co-locating storage with compute to eliminate the transfer bottleneck. The Copernicus Data Space Ecosystem (CDSE)[7] provides cloud-based access to the full Sentinel archive, with processing and analysis services running alongside the data. These platforms demonstrate the viability of cloud-native EO data access at scale, but their performance characteristics are specific to public cloud environments with effectively unlimited internal bandwidth. The question of how cloud-native formats perform under the bandwidth constraints typical of institutional on-premises deployments remains largely unexamined.

### 2.2 Object Storage Performance for Data-Intensive Workloads

S3-compatible object storage has become the default persistence layer for cloud-native data platforms. Gadban and Kunkel [8] analysed the performance of the S3 API for HPC workloads, demonstrating that storage scalability is closely tied to network behaviour and request concurrency, and that standard S3 implementations — including on-premises deployments such as MinIO — do not yet meet peak HPC performance expectations under conventional configurations. Astsatryan et al. [9] evaluated MinIO and LustreFS as converged cloud–HPC storage solutions, confirming that cloud-integrated object storage can offer a scalable alternative to traditional parallel file systems, but did not isolate the network variable from storage performance. A common limitation of existing object storage benchmarks is that they report throughput numbers without controlling for or characterising the network fabric; the result is that reported performance conflates storage I/O with network transfer, making it difficult to attribute bottlenecks or project performance under different network assumptions.

### 2.3 Data-Centre Networking and High-Bandwidth Fabrics

Kluge et al. [10] characterised sustained data movement across a wide-area 100 Gbps testbed, and their results, obtained on the first commercially available single-fibre 100 Gbps infrastructure, establish an empirical baseline for fabric characterisation. At leadership-computing scale, the Frontera system at TACC reflects the same principle — that interconnect bandwidth and topology are first-order determinants of usable system throughput rather than peripheral concerns — with HDR-100 InfiniBand to compute nodes and a fat-tree fabric designed for sustained parallel workloads [11]. Stewart et al. [12] documented the campus-bridging challenge: institutions producing terabytes of data per day face practical limits on data mobility when internal connectivity is limited to 1 or 10 Gbps, a constraint that 100 Gbps networking has the potential to resolve. However, these studies focus on general-purpose scientific data transfer rather than EO-specific workloads with their distinctive object size distributions and access patterns.

### 2.4 Distributed and Federated EO Infrastructure

The challenge of distributed EO data access has motivated several infrastructure initiatives. The European Open Science Cloud (EOSC) provides federated access to research data across European institutions. Federated data space initiatives address governance and interoperability but not the performance characteristics of the underlying infrastructure. At the other end of the spectrum, individual institutions operate isolated data archives with no mechanism for cross-site access or replication. Between these extremes — fully federated platforms and isolated local stores — there is a gap: practical models for loosely federated on-premises infrastructure, where each node is independently useful, operationally autonomous, and able to share data and services with partner nodes through standard protocols. This is the space the Atlantic Cloud occupies: not a centrally managed platform, but a network of independent nodes built on a common architecture, connected through S3-compatible replication and a shared API layer.

### 2.5 The Gap

The literature reveals a consistent absence of practical guidance. Cloud-native EO formats and platforms assume effectively unlimited internal bandwidth. Object storage benchmarks conflate network and storage performance. Network characterisation studies focus on general-purpose data transfer. Distributed infrastructure models address governance without characterising performance.

The consequence is practical: an institution planning an on-premises EO data node — whether standalone or as part of a distributed network — must decide how much to invest in network bandwidth, with no empirical basis for the decision. This paper provides that basis, reporting measured performance from the first operational node of the Atlantic Cloud under controlled conditions that isolate the key variables.

The absence matters more than a gap in the literature usually does, because the intuition that fills it makes the wrong predictions. An account in which throughput simply tracks bandwidth predicts that a tenfold reduction in per-server link rate costs a factor of ten; the measurements report 12–18 %, and the departure indicates a threshold rather than a slope. It predicts that object size matters; under throttled conditions it does not. It predicts that round-trip time governs what a remote partner can achieve; across 63 sites the class of network the path traverses matters more. And it predicts that a cluster reporting healthy under production monitoring is healthy; four latent faults at the AIR Data Centre were visible only under per-node measurement. Each is a place where the obvious account of network-bound storage is quantitatively wrong, and each is reported below.

## 3. The Atlantic Cloud

Earth observation is produced globally but consumed regionally. Institutions working on Atlantic coastal monitoring, maritime surveillance, land deformation, or ocean dynamics share overlapping areas of interest — from the Mid-Atlantic Ridge to the Brazilian coast, from West Africa to the Caribbean — but have historically operated in isolation, each maintaining its own data holdings with no mechanism for cross-site access or replication. Public cloud platforms offer shared infrastructure, but their regional coverage of the Atlantic basin is thin. Amazon Web Services (AWS) operates one region in Africa, at Cape Town, which is disabled by default and must be opted into, and one in South America, at São Paulo [13]; Google Cloud Platform (GCP) operates one African region, at Johannesburg, and two in South America, at São Paulo and Santiago [14]. Neither provider has a region in West Africa or anywhere in the island Atlantic. The effect is measurable from Terceira: in the campaign reported in §7, every cloud region within 100 ms is European, while the regions nearest the basin's southern and eastern rim sit at 224 ms (Johannesburg), 241 ms (São Paulo, AWS) and 253 ms (São Paulo, GCP). The resulting dependencies are difficult to reconcile with institutional autonomy, national sovereignty requirements, and the connectivity constraints of remote or island settings.

The Atlantic Cloud is a transatlantic network of on-premises data facilities initiated by the AIR Centre and its partner institutions to address this structural gap. The model rests on three principles.

First, each node is independently operational. A participating institution builds and operates its own infrastructure, serves its own users, and runs its own EO services. There is no dependency on a central platform or external provider for day-to-day operation. This matters in settings where connectivity is intermittent or where institutional mandates require local data custody.

Second, nodes share a common architecture. All nodes use S3-compatible object storage (MinIO), spatial metadata databases (PostGIS), and standard OGC APIs for external access. This commonality is not enforced centrally; it emerges from shared technical guidance and the practical advantages of interoperability. A service developed at one node can operate at another without re-engineering, and data stored at one node can be accessed from another using the same client tools.

Third, nodes can federate incrementally. MinIO provides built-in support for multi-site replication — bucket-level, asynchronous, and configurable per node — enabling cross-site data sharing without requiring a centralised coordination layer. An API gateway (services.aircentre.org) provides a single entry point for service discovery and access control across the network. Federation is not all-or-nothing: a new node can operate independently for months or years before enabling replication with partner nodes, and replication can be scoped to specific datasets or services rather than applied wholesale.

The Atlantic Cloud has one operational node. The AIR Data Centre on Terceira island (Azores, Portugal) is fully operational and the subject of the empirical evaluation in this paper. Partner institutions across the Atlantic basin — the North Carolina Institute for Climate Studies (NCICS/NOAA, USA), Instituto Federal Fluminense (IFF, Brazil), the Laboratório de Métodos Computacionais em Engenharia (LAMCE/UFRJ, Brazil), the National Space Research and Development Agency (NASRDA, Nigeria), the Laboratorio Nacional de Observación de la Tierra (LANOT/UNAM, Mexico), the University of Waterloo (Canada), and the Texas Advanced Computing Center (TACC, USA) have completed a first round of multi-site replication benchmarks, reported in §7, running a containerised MinIO instance supplied for the campaign rather than a deployed node, that characterise the federation primitive between partner sites and the AIR Data Centre; the campaign continues with additional partner sites under discussion. The architecture is designed to accept further nodes as partner institutions adopt it.

This paper characterises the performance of the AIR Data Centre node in detail. The architecture described in §4 is the reference design for all Atlantic Cloud nodes; the benchmarks in §5–7 establish what that architecture delivers under controlled conditions, providing the empirical basis that prospective node operators need to plan and justify their own deployments.

## 4. System Architecture

This section describes the architecture at two levels: the replicable pattern that any Atlantic Cloud node can follow, and the specific implementation at the AIR Data Centre that was benchmarked for this paper.

### 4.1 Architecture Pattern

The architecture follows a cloud-native, microservices design with three principal layers: data storage, metadata management, and external access. Each layer uses open-source, commodity components that can be sourced independently and deployed on standard server hardware.

**Data storage** is provided by MinIO, an open-source, S3-compatible object storage system. MinIO runs as a distributed cluster across multiple bare-metal nodes, using erasure coding for data protection and JBOD (Just a Bunch of Disks) drive configurations to maximise I/O throughput. The choice of MinIO is motivated by full S3 API compatibility (enabling direct use of cloud-native EO tools and formats), data protection through erasure coding (without overhead), horizontal scalability through the addition of server pools, and built-in support for multi-site replication, which allows federation between Atlantic Cloud nodes to be configured per institution and per bucket, preserving each node's operational autonomy. MinIO writes data and metadata together as objects, with no separate metadata server; this fully symmetric architecture means every node in the cluster can handle client requests, and load balancing is straightforward.

**Metadata management** is handled by a spatial database — PostGIS (the spatial extension for PostgreSQL) in the reference deployment — which maintains a catalogue of ingested EO data objects with their spatial footprints, temporal coverage, mission identifiers, and processing levels. The strict separation between payload storage (MinIO) and metadata (PostGIS) is a deliberate design decision: metadata queries are resolved without touching large data objects, reducing unnecessary data movement. Discovery queries — "what Sentinel-2 data covers this region in this time window?" — hit only the metadata database; the resulting object identifiers are then used to retrieve payloads directly from MinIO via the S3 API.

**External access** is provided through two standard interfaces. The S3 API provides direct object access for programmatic clients. The OGC API — Environmental Data Retrieval (EDR) [15] provides a query-based interface for discovery and subsetting, following the OGC API suite that is increasingly adopted by national and international EO data providers. An API gateway handles authentication, rate limiting, and request routing. These interfaces act as stable contracts: internal implementation details — which MinIO version, which database backend, which server hardware — can change without affecting clients.

**Network fabric** is treated as a first-class architectural component, not an assumed commodity. The internal network connecting storage and application nodes determines the upper bound on storage throughput, and therefore on every service the system can deliver. The architecture specifies a high-bandwidth, redundant fabric with bonded multi-port connections per node and dual switches for failover. The specific bandwidth is a deployment decision — this paper evaluates the consequences of that decision at 1, 10, and 100 Gbps.

### 4.2 AIR Data Centre Implementation

The AIR Data Centre is deployed in TERinov, the science and technology park on Terceira island, Azores. The facility is a purpose-built six-rack data centre currently housing 14 servers organised into five functional tiers, all interconnected by a common 100 GbE fabric. Table 1 summarises the deployment; the tiers directly involved in the benchmarks presented in §5–7 are described in full below, with the remaining tiers and the facility's network provisioning describe in (Fig. 1) and further detailed in Appendix A.

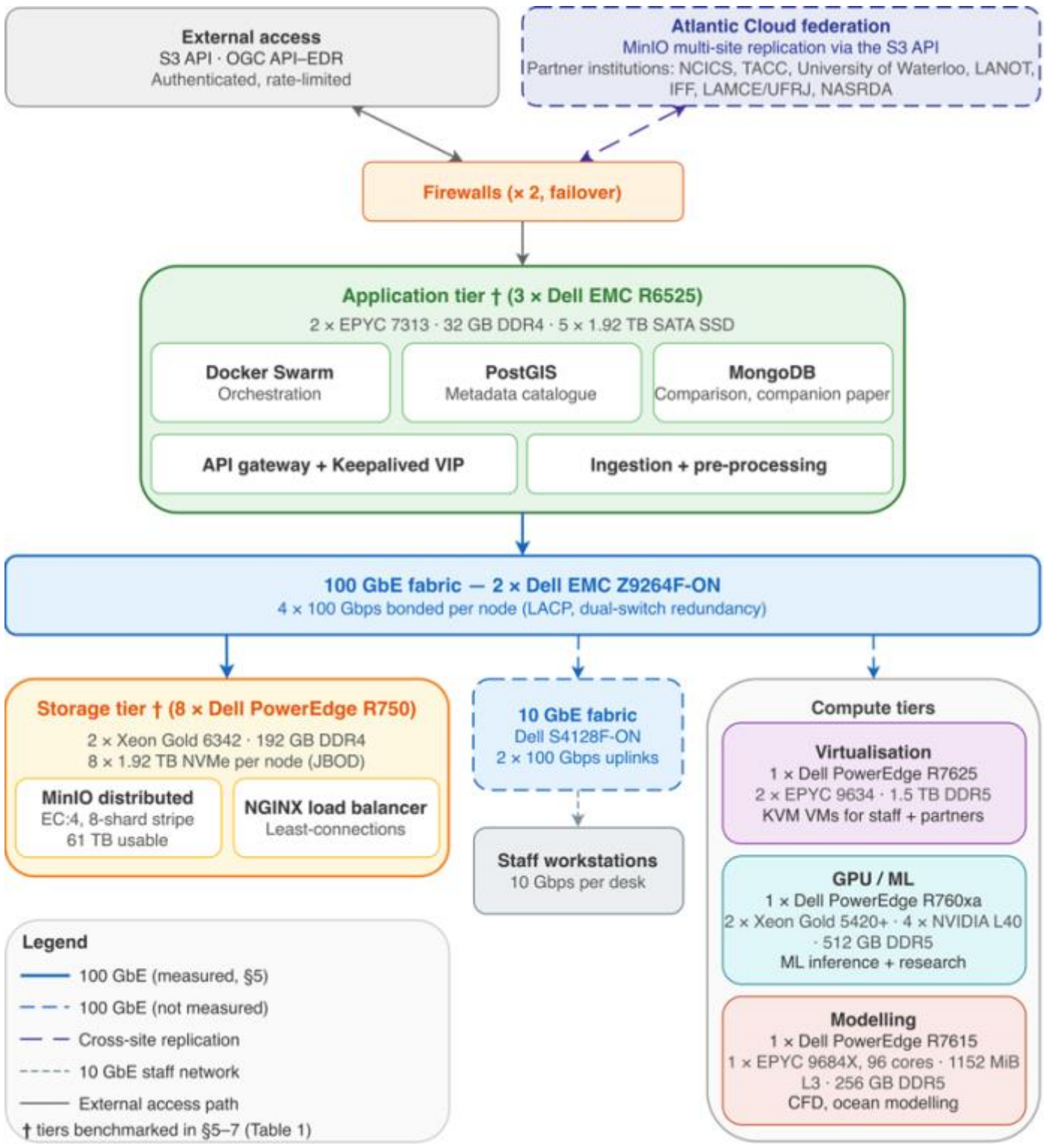

Fig. 1. AIR Data Centre overall system architecture: five server tiers, 100 GbE and 10 GbE fabrics, external access paths, and the Atlantic Cloud federation path.

Table 1. AIR Data Centre hardware deployment. All servers connect to the 100 GbE fabric via 2× Mellanox ConnectX-6 DX dual-port adapters (4× 100 Gbps bonded). Simultaneous multithreading is disabled on all nodes, so thread counts equal core counts. Tiers marked with † are directly involved in the benchmarks presented in §5–7.

| Tier | Nodes | Server | CPU (per node) | RAM | Storage (per node) | Role |
|---|---|---|---|---|---|---|
| Storage † | 8 | Dell PowerEdge R750 | 2× Xeon Gold 6342, 24C/24T, 2.8 GHz | 192 GB DDR4 | 2× 1.6 TB NVMe (OS) + 8× 1.92 TB NVMe (JBOD) | MinIO obj. storage |
| Application † | 3 | Dell EMC R6525 | 2× AMD EPYC 7313, 16C/16T, 3.0 GHz | 32 GB DDR4 | 5× 1.92 TB SATA SSD | Services, PostGIS, API |
| Virtualisation | 1 | Dell PowerEdge R7625 | 2× AMD EPYC 9634, 84C/84T, 2.25 GHz | 1.5 TB DDR5 | Per VM allocation | KVM VMs for staff + partners |
| GPU / ML | 1 | Dell PowerEdge R760xa | 2× Xeon Gold 5420+, 28C/28T, 2.0 GHz + 4× NVIDIA L40 | 512 GB DDR5 | Per workload | ML inference + research |
| Modelling | 1 | Dell PowerEdge R7615 | 1× AMD EPYC 9684X, 96C/96T, 2.55 GHz | 256 GB DDR5 | 4× 1.92 TB NVMe | CFD, ocean modelling |

**Network fabric.** Two Dell EMC Z9264F-ON switches provide the spine of the internal network. Every server in the facility — regardless of tier — connects to both switches via Mellanox ConnectX-6 DX dual-port 100 GbE QSFP56 adapters, two adapters per server, yielding four 100 Gbps ports per node. The four ports are bonded, providing approximately 400 Gbps of bonded bandwidth per node with full redundancy: losing any single switch or any two links per node leaves the remaining paths operational. Fig. 2 shows the dual-switch, four-port bonding and redundancy layout; full port allocation, cabling, and the staff network are given in Appendix A.

**Storage tier (8 nodes — benchmarked).** Eight Dell PowerEdge R750 servers, each equipped with dual Intel Xeon Gold 6342 processors (24 cores / 24 threads per socket, 2.8 GHz), 192 GB DDR4-3200 RAM, two 1.6 TB enterprise NVMe drives in RAID-1 for the operating system, and eight 1.92 TB enterprise NVMe drives in JBOD configuration for MinIO storage. Each chassis accommodates up to 24 NVMe drives, providing a growth path without node replacement. MinIO runs in distributed mode across all eight nodes, with 64 drives total. Erasure coding is configured as EC:4 — 4 data shards and 4 parity shards per 8-shard stripe — yielding 61 TB usable capacity from 122 TB raw. This configuration tolerates up to 4 simultaneous server failures and 32 drive failures across the cluster. Drives are formatted with XFS and presented to MinIO without hardware RAID. An NGINX load balancer distributes client requests across storage nodes using a least-connections algorithm.

**Application tier (3 nodes — benchmarked).** Three Dell EMC R6525 servers, each with dual AMD EPYC 7313 processors (16 cores / 16 threads per socket, 3.0 GHz), 32 GB DDR4-3200 RAM, and five 1.92 TB SATA SSDs. These nodes run Ubuntu Server and host all services via Docker Swarm, which provides container orchestration, load balancing, and automatic failover. A virtual IP managed by Keepalived provides automatic failover for the API gateway. Services deployed include the OGC API–EDR endpoint, the metadata ingestion pipeline, data pre-processing microservices, and the API access management system. PostGIS runs as a containerised service on this tier.

**Virtualisation tier (1 node).** One Dell PowerEdge R7625 with dual AMD EPYC 9634 processors (84 cores / 84 threads per socket, 2.25 GHz) and 1.5 TB DDR5-4800 RAM. This server hosts KVM virtual machines for AIR Centre staff and partner institutions, providing isolated development and analysis environments. The R7625 is characterised under a concurrent fan-out workload in §5.5 (Table 5b); its 24-channel DDR5-4800 memory subsystem sustains 94.5% of the 400 Gbps bond capacity, providing an empirical reference point for the bond-saturation regime that the R6525 application tier reaches once its memory channels are fully populated.

**Metadata database.** PostGIS runs as a containerised service on the application tier within the Docker Swarm cluster. The schema stores per-object metadata including mission, sensor, processing level, acquisition timestamp, and spatial footprint (as a PostGIS POLYGON geometry in WGS84). A second containerised instance of MongoDB runs alongside PostGIS to support a comparative evaluation of spatial-query and ingestion performance at EO catalogue scale, reported in a companion paper.

**External access.** The API gateway is accessible at services.aircentre.org. It provides authenticated access to the S3 API (for direct object retrieval), the OGC API–EDR endpoint (for query-based discovery and subsetting), and administrative interfaces for account management and rate limiting. This gateway also serves as the entry point for remote access by other Atlantic Cloud nodes.

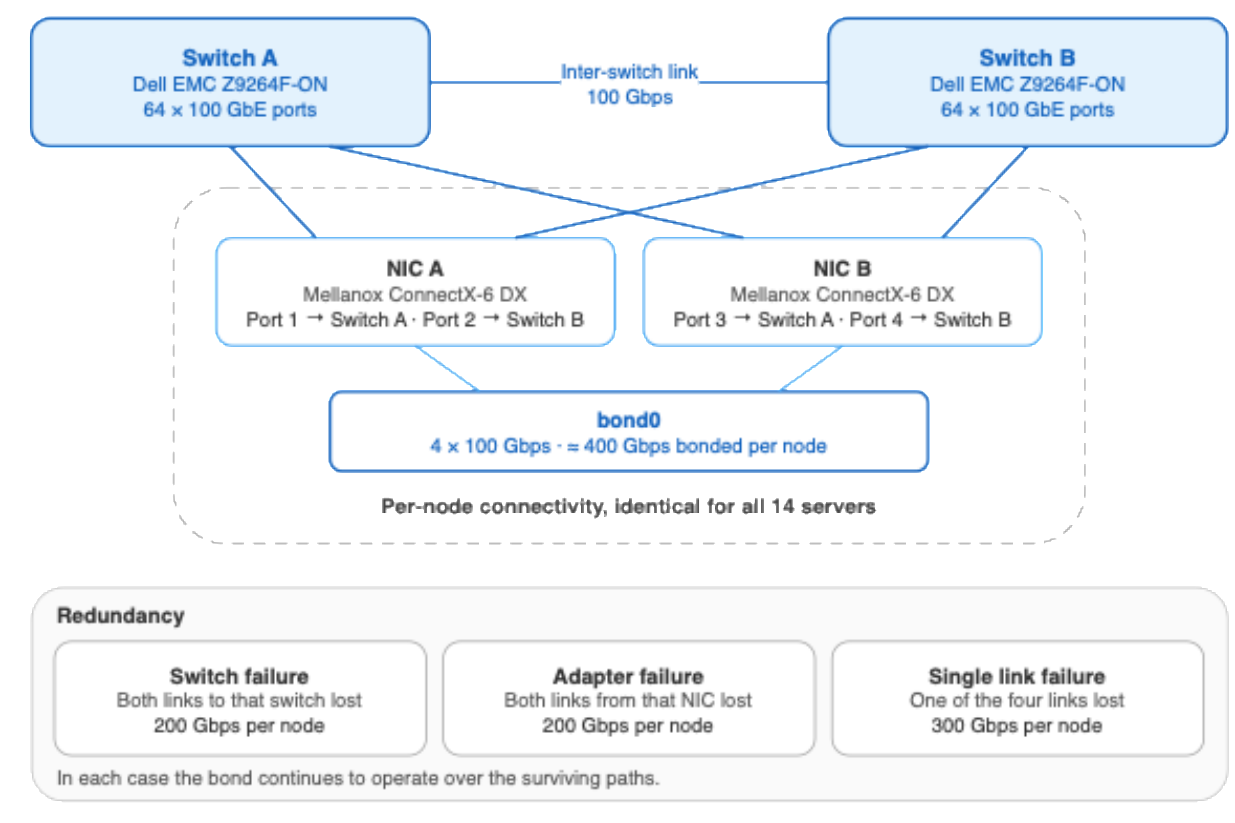

Fig. 2. Network topology: dual-switch, 4-port bonding configuration, and redundancy characteristics.

## 5. Network Fabric Characterisation

The network fabric determines the upper bound on storage throughput and therefore on every data access pattern the system can support. This section characterises the 100 GbE fabric described in §4 under sustained parallel load, evaluating throughput, stability, latency, and fault tolerance. The results establish the performance envelope within which the object storage evaluation of §6 operates.

### 5.1 Methodology

All throughput measurements use iperf3 v3.17.1, configured with 8 parallel TCP streams per session, BBR congestion control, 256 KB TCP window, 64 KB buffer length, and zero-copy mode disabled (-Z) for measurement consistency. Each session runs for 120 seconds. Three repetitions are recorded for every configuration, with system snapshots (CPU, memory, disk I/O, network counters) captured before each run. All measurements were taken with the application tier drained of production workloads (Docker Swarm services stopped) to eliminate interference. Raw iperf3 JSON output — including per-second interval data, per-stream breakdowns, CPU utilisation, TCP window sizes, and retransmit counters — is preserved for every session.

Throughput is measured in five modes, each isolating a different contention pattern:

The modes differ along two dimensions. **Per-session concurrency** is the number of flows on the fabric at any single instant during one measurement (1 for per-pair; 3 for storage-inbound; 8 for application-outbound; 24 for full-mesh; 16 for storage-to-storage). **Inter-configuration sequencing** is whether the multiple configurations within a mode are run simultaneously or one at a time: in per-pair, storage-inbound, and application-outbound the configurations run sequentially (one pair, one storage target, or one application node at a time, respectively); in full-mesh and storage-to-storage the entire fabric load is generated in a single concurrent run. Reported aggregate throughput in Table 2 sums the sent-side throughputs across all sessions within a mode; for the three sequential modes this is therefore a sum across non-concurrent measurements, not an instantaneous fabric load.

**Per-pair (sequential).** One iperf3 session at a time on the fabric: a single application node sends to a single storage node, and the 24 pairs are run sequentially. 24 pairs total (3 application × 8 storage), 3 repetitions each. This measures the peak throughput a single connection can sustain with no contention, establishing the baseline against which all other modes are compared.

**Application-outbound (1→8).** One application node drives 8 simultaneous sessions, one to each storage node. 3 configurations (one per application node), run sequentially with 3 repetitions each. This measures the aggregate outbound capacity of a single application node — the relevant scenario for a client downloading data from across the cluster.

**Storage-inbound (3→1).** All 3 application nodes send simultaneously to a single storage node. 8 configurations (one per storage node), run sequentially with 3 repetitions each. This measures the inbound capacity of a single storage node under multi-client load — the relevant scenario for parallel data ingestion.

**Full-mesh (3→8).** All 3 application nodes drive sessions to all 8 storage nodes simultaneously — 24 concurrent sessions. 3 repetitions. This is the operationally relevant configuration: the condition under which MinIO benchmarks in §6 will run.

**Storage-to-storage (4→4).** A supplementary test in which 4 storage nodes (st01–st04) drive sessions to the other 4 (st05–st08) — 16 concurrent sessions using the R750 hardware class on both sides. 3 repetitions. This isolates the switch fabric's capacity from the application node's memory bandwidth constraints, providing a reference for what the network can deliver when endpoint hardware is not the bottleneck.

Latency is measured using ping at 0.2-second intervals under three conditions: idle (no background traffic), 50% load (half the application nodes driving full-mesh traffic), and 100% load (all nodes active). Samples are collected from every application node to every storage node (216 measurement paths). Packet loss is assessed by comparing interface error counters (ip -s link) before and after sustained full-mesh load across all 11 nodes.

Fault tolerance is observed during the initial deployment validation phase, under sustained full-mesh load, across five progressive failure scenarios spanning the redundancy envelope from single-link to multi-link/switch loss. The scenarios and the observed behaviour are reported in Table 7 (§5.8); systematic quantification of recovery times with full statistical repetitions is planned for the next scheduled maintenance window and reported separately.

### 5.2 Throughput

Table 2. Aggregate throughput by measurement mode (mean ± std, 3 repetitions).

| Mode | Sessions | Aggregate (Gbps) | Per-pair mean (Gbps) | Retransmits |
|---|---|---|---|---|
| Per-pair (sequential) | 1 | 2,570.3 ± 11.5 | 107.1 | 13 |
| Storage-inbound (3→1) | 3 | 2,458.0 ± 34.1 | 102.4 | 19 |
| App-outbound (1→8) | 8 | 522.9 ± 4.3 | — | 21 |
| Full-mesh (3→8) | 24 | 529.0 ± 14.0 | 22.0 | 20 |
| St-to-st (4→4) | 16 | 1,378.6 ± 31.1 | 86.2 | 199 |

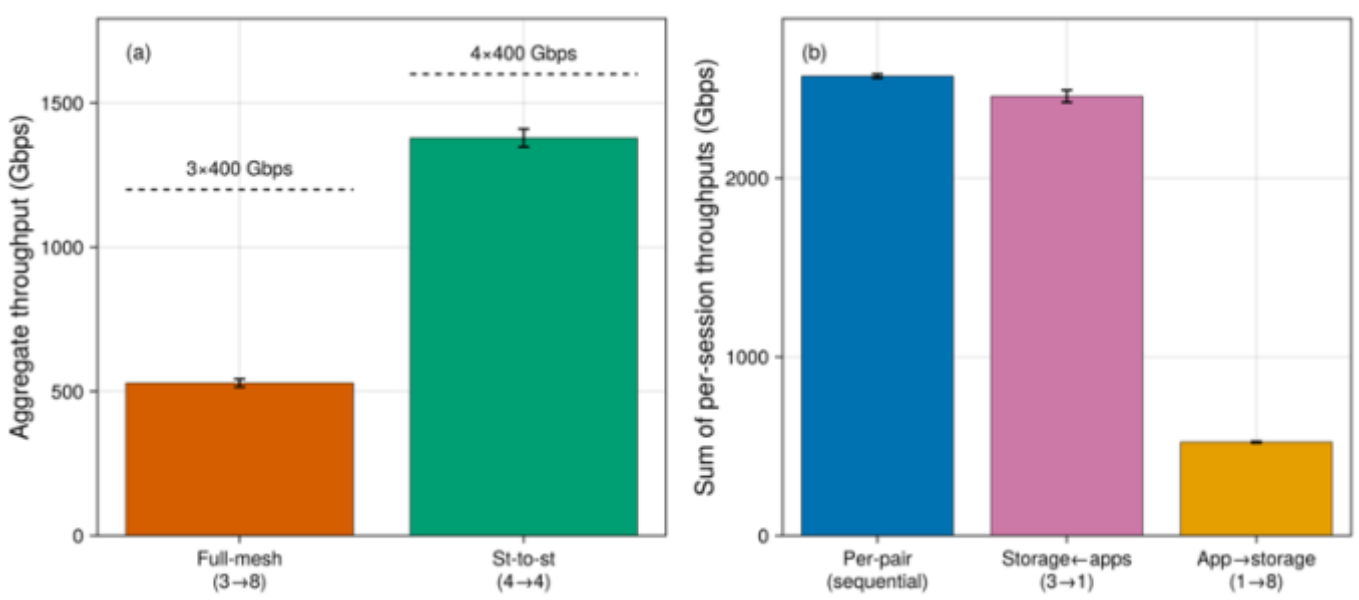

Fig. 3. Aggregate throughput by measurement mode, in two panels. (a) The two modes whose aggregate is an instantaneous fabric-wide load: full-mesh (3→8) and storage-to-storage (4→4). Dashed reference lines drawn over each bar mark the sender-side NIC-bond ceiling for that mode: 3 × 400 Gbps for full-mesh and 4 × 400 Gbps for storage-to-storage. (b) The three modes whose aggregate is a sum across non-concurrent measurements: per-pair (sequential), storage-inbound (3→1), and application-outbound (1→8). A fabric-wide ceiling does not apply to these aggregations (see §5.1), so no reference lines are drawn. Bars show the mean across three repetitions; whiskers show ±1 standard deviation.

The aggregate-by-mode results in Table 2 are visualised in Fig. 3. The per-pair baseline confirms that the fabric delivers over 100 Gbps per connection with no contention, consistent across all 24 pairs (range: 98–113 Gbps). Application-outbound runs each application node in isolation against all eight storage targets (Table 2: 522.9 Gbps summed across the three nodes, mean 174 Gbps per node). The per-node outbound figures (srv01: 167.7 Gbps; srv02: 167.0 Gbps; srv03: 188.2 Gbps) are within 2% of each app server's mean contribution under full-mesh contention (srv01: 168.7; srv02: 168.7; srv03: 191.7 Gbps). A single application server driving the fabric in isolation is no faster than under contention: the application endpoints are the limit, not fabric capacity or cross-node interference. The cause is memory bandwidth on the R6525 application tier — a known constraint at procurement and the planned next upgrade. The st04 node, which initially underperformed at approximately 84 Gbps due to a BIOS configuration error (CPU frequency boost disabled), was corrected and re-measured; all 8 storage nodes are now within the same performance band (Fig. 4).

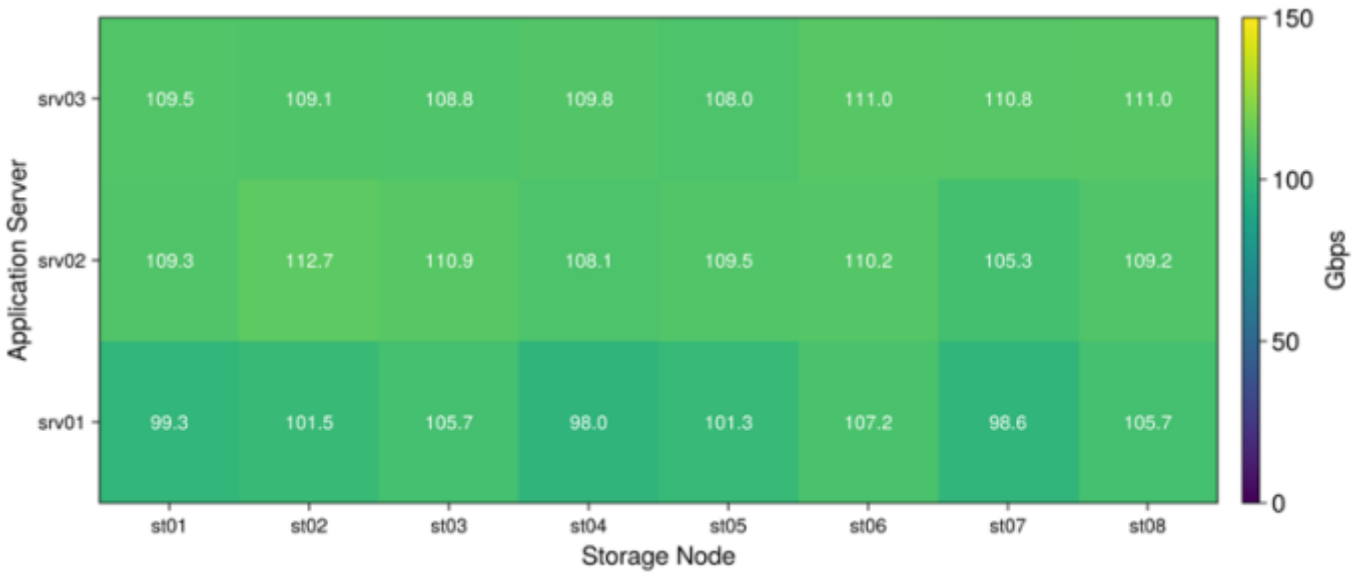

Fig. 4. Per-pair baseline throughput across the 24 application–storage pairs, with no contention. Each cell shows the mean Gbps over the measurement window for one (application, storage) pair.

Under full-mesh contention, aggregate throughput drops to 529 Gbps — 44% of the 1,200 Gbps theoretical maximum (3 application nodes × 400 Gbps bond each). Degradation per pair averages 79.5% (range: 66.7–88.3%) and is uniformly distributed across all 24 pairs, indicating that the bottleneck is not at the switch fabric or at individual links, but at the application node endpoints.

The storage-to-storage test confirms this diagnosis. When R750 storage nodes serve as both clients and servers, the same fabric sustains 1,378.6 Gbps under symmetric 4×4 contention — 86.2% of the 1,600 Gbps theoretical maximum (4 nodes × 400 Gbps). The R750 servers, with twelve-channel DDR4-3200 (six per socket) memory at 192 GB per node, push 320–359 Gbps per node (80–90% of their 400 Gbps bond) from the client-side view, and 339–352 Gbps per node from the server-side view. The R6525 application servers, with two-channel DDR4-3200 (one per socket) memory at 32 GB per node, are limited to mean 176 Gbps per node (169–192 range, 44% of bond capacity). The difference is attributable to memory bandwidth: at 100 Gbps per stream and 8 parallel streams, data movement between the NIC and application memory saturates the R6525's memory bus before the network bond or the switch fabric is fully utilised.

This asymmetry between the storage and application tiers is not accidental — it reflects a deliberate procurement strategy driven by a constrained initial budget. The AIR Data Centre was designed around a principle: invest upfront in components that are difficult and disruptive to change once operational (network fabric, cabling, switch architecture), and start conservatively on components where incremental upgrades carry low penalty (server memory, storage capacity, compute nodes). The R6525 application nodes were procured with a single memory channel populated per socket, leaving the remaining channels available for incremental expansion. Fully populating the memory channels of the existing R6525 nodes — a planned, low-disruption intervention requiring no change to network, racks, or software — is expected to substantially raise per-node memory bandwidth, bringing the application tier closer to the regime where the storage-to-storage result (1,378.6 Gbps under contention, 86.2% of theoretical) demonstrates the fabric can sustain. Network infrastructure — switches, optics, cabling runs between racks — is painful to replace in a running facility; adding memory or replacing application servers is routine. The benchmarks validate this strategy: the 100 GbE fabric, which represents the upfront investment, is not the constraint. The R6525 application nodes, which represent the deferred investment, are — and upgrading them to higher memory bandwidth (e.g., R750-class or DDR5 servers) requires no change to the network. The st-to-st result demonstrates what the fabric already supports: an institution that starts with the same architecture and progressively upgrades its application tier would see aggregate throughput scale from the current 529 Gbps toward the 1,379 Gbps the fabric can sustain.

This finding has direct implications for institutions planning on-premises EO deployments. The 44% fabric utilisation observed under full-mesh is specific to the R6525 application tier, not a property of the network. An institution deploying the same architecture on uniform R750-class hardware would expect aggregate throughput closer to 85% of theoretical. In practical terms, network bandwidth is the investment with the highest long-term return — it determines the ceiling — while endpoint memory bandwidth (memory channel population, DIMM speed, and NUMA-aware NIC placement; see §5.5) determines how much of that ceiling is realised at any point in time, and can be upgraded incrementally as workloads and budgets grow.

### 5.3 Throughput Stability

Per-second throughput intervals (120 data points per session) reveal substantial variation in stability across modes. The coefficient of variation (CV) of per-second aggregate throughput is reported in Table 3.

Table 3. Per-second throughput stability (CV%) by mode.

| Mode | Mean CV (%) | Worst case (%) |
|---|---|---|
| St-to-st (4→4) | 2.75 | 4.42 |
| Storage-inbound (3→1) | 4.92 | 13.37 |
| Per-pair (sequential) | 5.28 | 11.38 |
| Full-mesh (3→8) | 14.06 | 44.47 |
| App-outbound (1→8) | 14.99 | 44.56 |

The storage-to-storage mode is the most stable, with no pair exceeding 4.5% CV. This confirms that the fabric delivers predictable throughput when endpoint hardware is not constrained. In contrast, the full-mesh and application-outbound modes exhibit bursty behaviour, with some sessions reaching 44% CV — driven by BBR congestion control adjusting bandwidth estimates as competing flows negotiate shared memory bandwidth at the application node. The full-mesh time series (Fig. 5, third panel) shows a characteristic regime change at approximately 60 seconds, consistent with BBR's probing and draining phases settling after initial convergence.

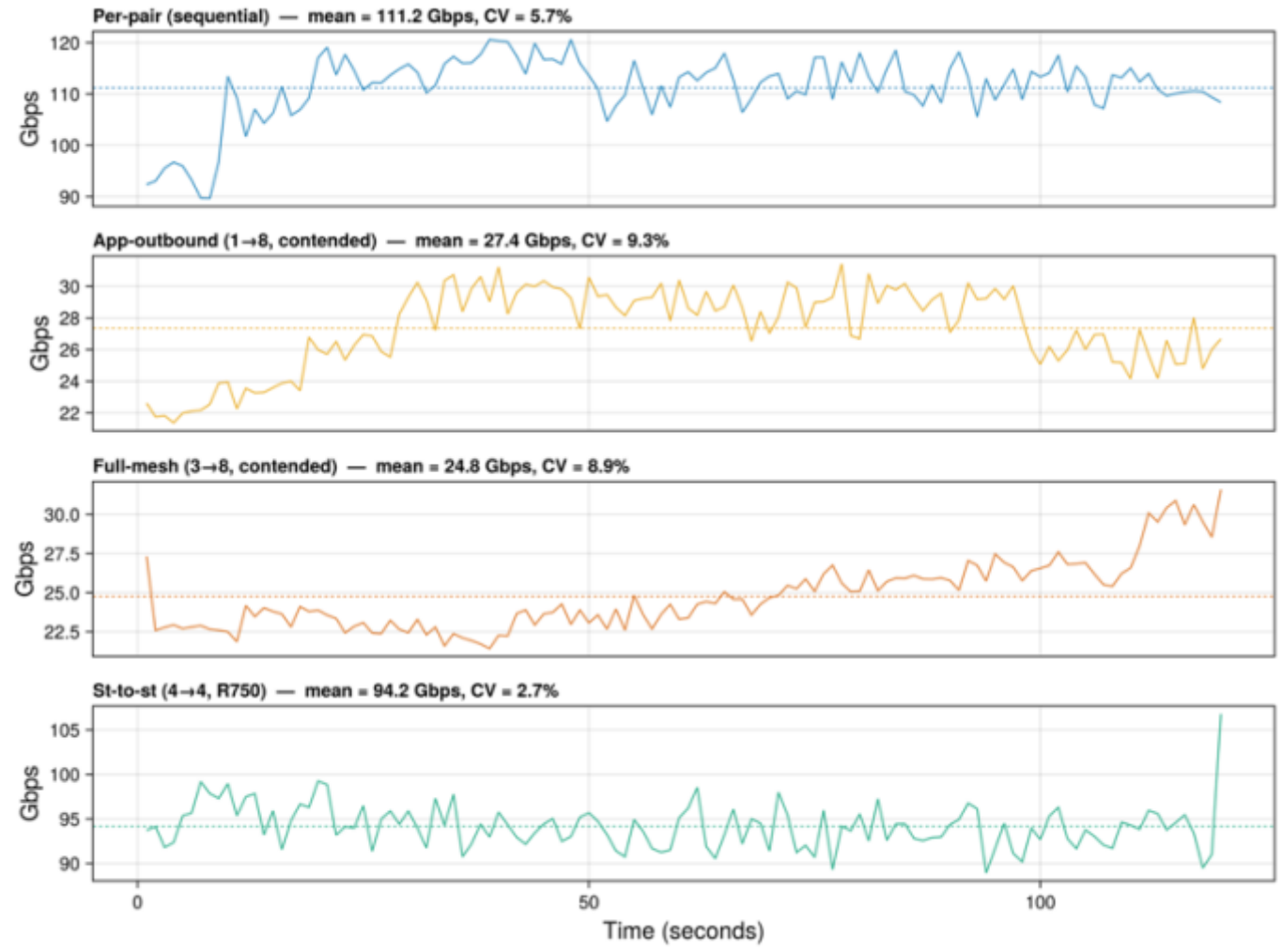

Fig. 5. Per-second throughput time series for one representative application–storage pair across four measurement modes. Each panel covers a 120-second iperf3 session at one-second resolution; dashed lines mark the per-session mean.

For EO workloads, stability matters because cloud-native access patterns — parallel Zarr chunk reads, COG tile streaming — are latency-sensitive at the individual request level. A 14% CV in aggregate throughput translates to unpredictable per-request latency, which is relevant for interactive analysis workflows. The st-to-st result (2.75% CV) indicates that upgrading the application tier to higher memory bandwidth hardware would improve not only aggregate throughput but also its consistency — reducing the burstiness that EO workflows are sensitive to.

### 5.4 Stream Balance and LACP Behaviour

Each iperf3 session uses 8 parallel TCP streams to saturate the bonded link. The distribution of throughput across those 8 streams reveals how effectively the LACP bond distributes traffic. Table 4 reports the coefficient of variation of per-stream throughput.

Table 4. Stream balance (CV% across 8 streams) by mode.

| Mode | Mean stream CV (%) | Worst case (%) |
|---|---|---|
| Per-pair (sequential) | 9.59 | 20.97 |
| Storage-inbound (3→1) | 13.19 | 37.68 |
| St-to-st (4→4) | 22.84 | 42.22 |
| App-outbound (1→8) | 76.55 | 132.84 |
| Full-mesh (3→8) | 93.28 | 162.69 |

Under no contention (per-pair), the 8 streams are well balanced: 9.6% CV, with each stream carrying approximately 13 Gbps of the 107 Gbps total. Under full-mesh contention, balance collapses to 93.3% CV — some streams carry 5–6 Gbps while others receive less than 1 Gbps (Fig. 6). This is a consequence of LACP hash distribution: the bond hashes flows to physical links based on source/destination IP and port tuples, and with 24 concurrent sessions sharing the same 4 physical links, hash collisions concentrate traffic on a subset of links. The result is that aggregate throughput is limited not by the sum of link capacities, but by the most loaded link.

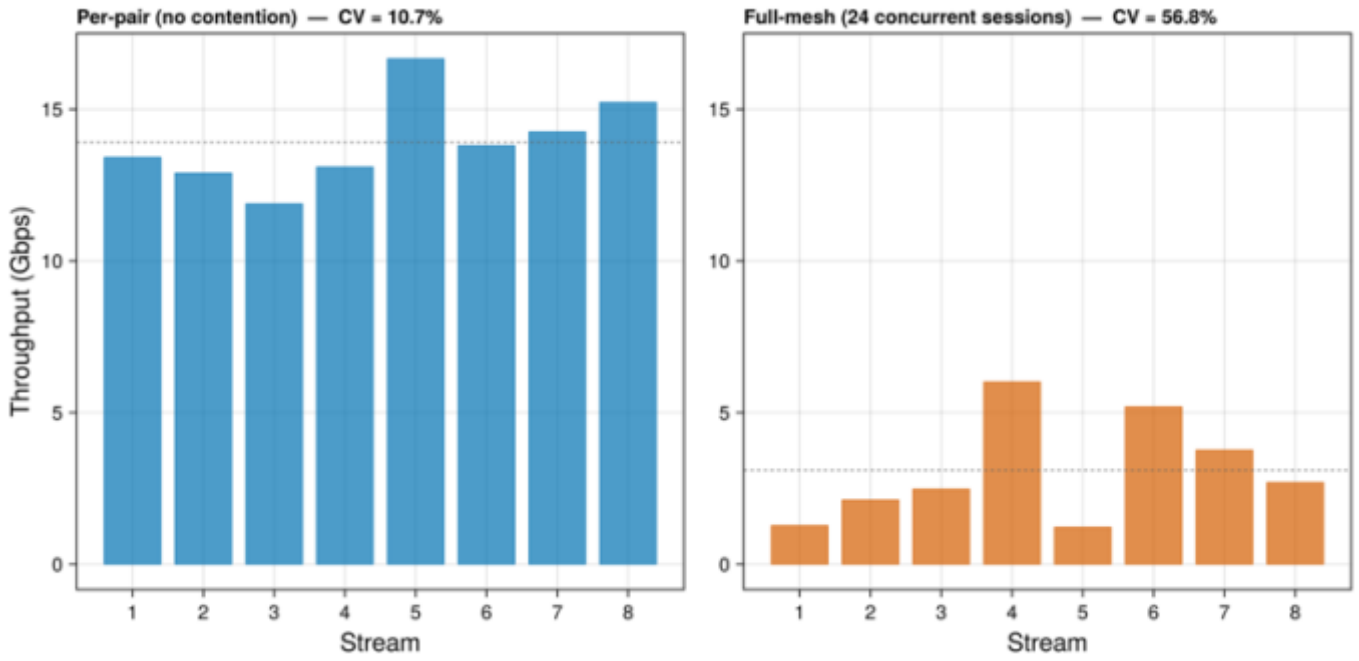

Fig. 6. Per-stream throughput for one representative application–storage pair under per-pair (no contention) and full-mesh (24 concurrent sessions) conditions. Each bar shows the aggregate throughput of one of the eight TCP streams used to saturate the bonded link. Per-panel CV annotations report stream balance for the representative session shown; mean per-session CV across all pairs in each mode is reported in Table 4.

This is a known limitation of LACP bonding under high-concurrency workloads and has been documented in the context of HPC fabrics. The practical consequence for EO data access is that the effective bandwidth available to any single client session under parallel load is less predictable than the aggregate numbers suggest. The storage-to-storage mode, with its 22.8% stream CV, shows that higher-bandwidth endpoints mitigate the effect by reducing the number of flows competing for each physical link.

### 5.5 CPU Utilisation, Memory Bandwidth, and NUMA Topology

CPU utilisation during throughput tests provides evidence that memory bandwidth, not CPU capacity or network link speed, is the binding constraint on the application tier. Table 5 reports host-side CPU utilisation aggregated by mode.

Table 5. Host CPU utilisation (%) by mode (host = sender / client side).

| Mode | Mean CPU (%) | Std (%) | Throughput per node (Gbps) |
|---|---|---|---|
| Per-pair (sequential) | 246.9 | 29.2 | ~857 (sequential sum) |
| Storage-inbound (3→1) | 228.8 | 35.6 | ~820 (sequential sum) |
| St-to-st (4→4) | 54.7 | 7.7 | ~345 |
| App-outbound (1→8) | 38.3 | 15.3 | ~174 |
| Full-mesh (3→8) | 39.1 | 15.8 | ~176 |

CPU utilisation percentages above 100% reflect multi-core usage (e.g., 247% ≈ 2.5 fully utilised cores). Under per-pair sequential load, the R6525 application nodes saturate at approximately 247% CPU while pushing ~857 Gbps across their 8 targets sequentially — the CPU is fully engaged in data movement. Under full-mesh contention, the same nodes drop to 39% CPU while pushing only ~176 Gbps. The CPU is idle, waiting on memory bandwidth. The R750 storage nodes in the st-to-st test use 55% CPU to push 345 Gbps — more throughput at lower CPU utilisation, consistent with their higher memory bandwidth.

This pattern — high CPU at high throughput (sequential), low CPU at low throughput (contended) — is diagnostic of a memory bandwidth bottleneck rather than a CPU or network bottleneck. The host interconnect is not a third candidate: the four 100 GbE ports on each of these node classes are carried by two PCIe Gen4 ×16 adapters, whose combined capacity exceeds the 400 Gbps bond, and per-node throughput under contention is below what a single adapter alone can carry. Inspection of the NUMA topology on each node class explains why.

The R6525 application servers are configured with NPS=4 (Nodes Per Socket = 4), which partitions each AMD EPYC socket into four NUMA domains — eight NUMA nodes across the two-socket system. However, only one DIMM is populated per socket (16 GB each, 32 GB total), meaning that only 2 of the 8 NUMA nodes have local memory. The remaining 6 NUMA nodes — hosting 24 of the 32 available cores — must access memory remotely. Within the same socket, cross-node access incurs a 20% latency penalty (NUMA distance 12 vs 10); cross-socket access incurs a 220% penalty (distance 32). The theoretical memory bandwidth of this configuration is 51.2 GB/s (two DDR4-3200 channels), but the effective bandwidth under network load is substantially lower because most cores contend for remote memory access. This NUMA fragmentation, combined with the limited channel count, explains why the R6525 nodes achieve only 44% of their 400 Gbps bond capacity (mean 176 Gbps, 169–192 range, per node) under parallel load.

The R750 storage servers present a contrasting topology. Each socket has six populated DDR4-3200 channels (12 channels total, 192 GB), yielding a theoretical memory bandwidth of 307 GB/s. The NUMA topology is a standard two-node model (one per socket), with all cores on each socket having local memory. Cross-socket access incurs a moderate 30–40% penalty, but with balanced memory distribution the system rarely forces cross-socket traffic for local workloads. This explains why the R750 nodes achieve 86% of bond capacity (339–352 Gbps per node) in the storage-to-storage test — memory bandwidth comfortably exceeds the 50 GB/s required to saturate the 400 Gbps bond.

Table 5a summarises the memory architecture across the two benchmarked node classes and the modelling node (R7615) (full specification in Appendix A), which uses single-socket DDR5-4800 with four channels and no NUMA penalty.

Table 5a. Memory architecture and projected network saturation by node class. Memory bandwidth figures are theoretical, derived from channel count and per-channel rate. The saturation column is projected from memory topology and channel count for every node class rather than measured; measured per-node throughput is reported in Table 5 and in the text. Direct per-pair benchmarks of the R7615 are in progress.

| Node class | DIMM config | Channels | Mem BW (GB/s) | NUMA nodes | Can saturate 400G? (projected) |
|---|---|---|---|---|---|
| R6525 (application) | 2 × 16 GB DDR4-3200 | 2 (1/socket) | ~51 | 8 (NPS=4) | No (80–170 Gbps) |
| R750 (storage) | 12 × 16 GB DDR4-3200 | 12 (6/socket) | ~307 | 2 | Yes (672+ Gbps) |
| R7615 (modelling) | 4 × 64 GB DDR5-4800 | 4 (single socket) | ~154 | 1 | Yes (640+ Gbps) |

Table 5b. Concurrent fan-out from the R7625 (r04-rds02) to all eight storage nodes. Per-target session figures are means across the eight destinations; aggregate figures sum across destinations per rep, then mean across reps.

| Attribute | R7625 (r04-rds02) |
|---|---|
| CPU | 2 × AMD EPYC 9634 (84 cores / 84 threads per socket) |
| Sockets / cores total | 2 / 168 |
| Memory | 1,536 GiB (24 × 64 GiB DDR5-4800) |
| Channels populated | 24 (12 per socket) |
| Theoretical memory bandwidth | 921.6 GB/s |
| Bond capacity (sender) | 400 Gbps |
| Aggregate throughput | 377.8 ± 5.9 Gbps |
| % of bond capacity | 94.5% |
| Per-target throughput (mean across 8 destinations) | 47.2 Gbps |
| Per-target range | 43.7 – 49.0 Gbps |
| Target CV | 3.92% |
| Retransmits (campaign mean per rep) | 3 |

In a concurrent fan-out experiment from the R7625 to all eight storage nodes, the sender sustained 377.8 Gbps aggregate — 94.5% of the 400 Gbps bond capacity — distributed evenly across the eight destinations (per-target CV 3.92%; range 43.7–49.0 Gbps). At 921.6 GB/s of theoretical memory bandwidth across 24 populated DDR5-4800 channels, this tier sits an order of magnitude above the ≈50 GB/s required to saturate the bond, and the result confirms that memory bandwidth ceases to be the binding constraint when the memory subsystem is provisioned generously relative to the network. The retransmit count (3 per rep) is consistent with the low-loss regime reported in §5.7.

The application tier's 44% bond utilisation reflects the deliberately under-provisioned memory configuration of the R6525 nodes, not a limitation of the chassis. The platform supports 16 memory channels (8 per socket); only 2 are currently populated. Full population at DDR4-3200 would raise the theoretical memory bandwidth from ~51 GB/s to ~410 GB/s — well past the ~50 GB/s the bond requires to saturate — and would also place sufficient bandwidth on every NUMA node (51.2 GB/s per node under NPS=4), eliminating the cross-NUMA fragmentation that bottlenecks the current

## 5.6 Latency

Round-trip latency between application and storage nodes is measured under three load conditions. Table 6 reports the results.

Table 6. Round-trip latency (ms) between application and storage nodes.

| Condition | Mean (ms) | p95 (ms) | Max (ms)† |
|---|---|---|---|
| Idle | 0.053 | 0.075 | 0.84 |
| 50% load | 0.047 | 0.081 | 4.67 |
| 100% load | 0.072 | 0.173 | 2.59 |

† Worst observed RTT across all 24 application–storage pairs and 3 repetitions per pair under each condition; not a percentile.

Mean and p95 round-trip latencies stay sub-0.1 ms under all three conditions, with mean values between 0.05 and 0.07 ms. Tail latency, reported as the worst observed RTT across the 24 application–storage pairs, rises with load: 0.84 ms idle, 2.59 ms at 100% load, and 4.67 ms at 50% load (a single rare excursion on one pair). These tails are consistent with brief queueing events on individual links rather than sustained queueing across the fabric: if the throughput degradation described in §5.2 were driven by switch-level queueing, mean and p95 latency would also rise materially under load, which they do not. The degradation reported in §5.2 is therefore consistent with endpoint memory bandwidth limits rather than fabric congestion. For EO access patterns, sub-millisecond mean fabric latency means the dominant latency contribution in any data retrieval operation will be the storage I/O and application-layer processing, not the network.

### 5.7 Packet Loss

Interface error counters were compared before and after sustained full-mesh load across all 11 nodes. A total of 3 errors were recorded across 15.88 billion packets transferred — an error rate of $1.9 \times 10^{-10}$. Effectively zero packet loss under sustained parallel load, confirming that the fabric operates well within its engineered capacity.

### 5.8 Fault Tolerance

Systematic fault-tolerance characterisation with full statistical repetitions is planned for the next scheduled maintenance window and will be reported separately; this section reports the qualitative behaviour observed during the initial deployment validation phase.

Table 7. Failover behaviour under progressive link/switch failure.

| Scenario | Description |
|---|---|
| 1 | Single link failure on a storage node (1 port to one switch): LACP renegotiates; throughput recovers automatically at reduced capacity. |
| 2 | Symmetric dual link failure (one link per switch): LACP renegotiates via surviving ports; throughput recovers automatically at reduced capacity. |
| 3 | Asymmetric dual link failure (two links to the same switch): recovery is markedly slower than scenarios 1 and 2 but completes at reduced capacity. |
| 4 | Three links disconnected (one port remaining): node remains reachable via the surviving port but operates at 1/4 bond capacity. |
| 5 | Full switch disconnect (both ports to one switch): node is isolated when combined with any failure on the other switch; the 4-port bond tolerates up to 2 link failures per node distributed across both switches. |

Scenarios 1–3 demonstrate that the dual-switch, 4-port bonding design recovers automatically from single-point link failures. Recovery is dominated by LACP bond renegotiation and TCP session re-establishment. Scenario 5 (full switch disconnect) is the most operationally significant: the affected node loses half its bonded paths but remains reachable through the surviving switch, with throughput reduced proportionally. MinIO's distributed erasure coding architecture tolerates the temporary loss of a single storage node during recovery without data unavailability.

Scenarios 4 and 5 sit at the boundary of the designed redundancy envelope. Under Scenario 5 (full switch disconnect, both ports to one switch lost) the node continues to operate at half bond capacity through the surviving switch; an additional concurrent link failure on the surviving switch would isolate the node. Under Scenario 4 (three of four links lost) the node continues to operate at quarter bond capacity through the one remaining link; any further link loss would isolate it. The 4-port bonding configuration therefore tolerates any single-component fabric failure and any two-component failure that leaves at least one path open per node; correlated multi-failure events that exhaust all paths to a node — three or more concurrent link losses on that node, distributed across both switches — fall outside the envelope.

### 5.9 Summary

The network fabric sustains 1,379 Gbps aggregate throughput under symmetric R750 load (86% of theoretical), confirming that the switch fabric and cabling are not limiting. The operational 3→8 configuration achieves 529 Gbps, bounded by the R6525 application nodes' memory bandwidth, not by network capacity. Latency is sub-millisecond under all load conditions, packet loss is negligible, and the fabric recovers automatically from single-component failures. These results establish the performance envelope for the object storage evaluation in §6: MinIO throughput cannot exceed what the fabric and application tier together deliver, and the throttled baselines in §6 will show what happens when the network — not the endpoint — becomes the constraint.

## 6. Object Storage and Data Access Evaluation

This section evaluates MinIO object storage performance under the network fabric characterised in §5. The experiments use four object sizes representative of real EO data products, measure throughput under single-client and distributed-client configurations, characterise mixed read/write behaviour, and — critically — repeat the measurements at three of those sizes under throttled network conditions at 10 Gbps and 1 Gbps to isolate the effect of network bandwidth on storage throughput.

### 6.1 Methodology

All storage benchmarks use warp (MinIO's open-source benchmark tool), configured with parallel client processes generating sustained S3 PUT and GET workloads against the 8-node MinIO cluster described in §4.2. Each experiment runs for 120 seconds with 3 repetitions. System snapshots are captured before each run. Raw warp JSON output — including per-second segment data, per-host throughput breakdowns, and error counts — is preserved for every session.

**Object sizes.** Four sizes are tested, each representing a real EO data product and its typical access pattern:

**4 MiB** — Zarr chunk. The unit of random read access for cloud-native multi-dimensional datasets. Small enough that per-request overhead dominates throughput.

**64 MiB** — COG tile. The typical internal tile size for Cloud-Optimised GeoTIFF, accessed via HTTP range requests during spatial subsetting.

**512 MiB** — Sentinel-2 L2A granule. A representative bulk transfer unit for optical EO products.

**2 GiB** — Full SAR scene. The largest common EO data object, representing a complete Sentinel-1 IW mode acquisition.

**Experiments.** Four experiment types are conducted:

**Single-client (2.1).** One warp client (32 concurrent connections) accesses the cluster through the NGINX load balancer VIP. This represents the performance available to a single user or service — the baseline that any API client would experience.

**Distributed multi-client (2.2).** Three warp clients — one orchestrator on srv01 and two warp workers on srv02 and srv03 — drive 32 concurrent connections each, for 96 connections in total. All three clients access the eight storage nodes directly, bypassing the load-balancer VIP to avoid a single-point bottleneck at the gateway. This represents the aggregate capacity available under parallel access — the scenario for multi-user or multi-service workloads.

**Mixed read/write (2.3).** All three warp clients (orchestrator on srv01 plus workers on srv02 and srv03) run simultaneous GET and PUT operations (50/50 split by request count) at two sizes (4 MiB and 512 MiB). This represents operational conditions where data ingestion and retrieval occur concurrently — the normal state of an active EO data facility.

**Network-throttled baselines (2.4).** The distributed benchmark is repeated with Linux traffic control (tc) applied independently on each of the three application nodes — storage nodes are unmodified. Each application node is throttled to 10 Gbps and 1 Gbps in turn, using a token-bucket filter (tbf) on the bond0 egress path and the same filter on an Intermediate Functional Block (ifb0) interface that receives bond0 ingress traffic. This bidirectional shaping is necessary because tc tbf applied to bond0 alone shapes only egress traffic, leaving GET ingress at line rate; the IFB redirect closes that gap. The throttle is verified before each campaign with bidirectional iperf3 and the script aborts if reverse-direction throughput exceeds 120% of the target.

Because the throttle is applied per application node, the cluster's aggregate ceiling at a stated rate is approximately three times that rate (≈3 Gbps at the 1 Gbps configuration, ≈30 Gbps at the 10 Gbps configuration). Throughout this paper, the stated bandwidth labels refer to the per-application-node line rate; aggregate measurements should be read against the corresponding three-node ceiling. The number of concurrent connections per warp client is adjusted to fit the throttled rate (4 at 1 Gbps, 32 at 10 Gbps); 32 concurrent connections per client at 1 Gbps would exceed MinIO's multipart timeout for the 512 MiB object size. With the per-rate concurrency setting, all other configuration is held constant: same cluster, same drives, same clients, only the network rate changes. The 1 Gbps baseline corresponds to the commercial networking proposal rejected during the AIR Data Centre's procurement; 10 Gbps represents the upper end of conventional institutional networking. At the throttled rates the campaign covers three object sizes (4 MiB, 64 MiB, 512 MiB); 2 GiB is omitted because the 1 Gbps and 100 Gbps contrast is already visible at the smaller sizes.

**Measurement conditions.** The initial benchmarks (single-client, PUT, mixed, and throttled experiments) were conducted with production workloads drained from the application tier (Docker Swarm services stopped) to eliminate interference. Two configuration issues identified during distributed benchmarking — a TCP SACK setting on storage node st06 and TCP congestion-control drift across four of the eight storage nodes — were remediated before the distributed GET data reported in Table 9 was collected. The §8.4 discussion treats both as diagnostic findings.

### 6.2 Single-Client Performance

Table 8 reports single-client throughput across all four object sizes.

Table 8. Single-client MinIO throughput (32 concurrent connections via VIP).

| Size | PUT (MiB/s) | GET (MiB/s) | PUT (Gbps) | GET (Gbps) |
|---|---|---|---|---|
| 4 MiB | 5,576 ± 121 | 9,966 ± 296 | 46.8 | 83.6 |
| 64 MiB | 7,099 ± 223 | 10,019 ± 895 | 59.6 | 84.0 |
| 512 MiB | 7,369 ± 116 | 10,444 ± 212 | 61.8 | 87.6 |
| 2 GiB | 7,288 ± 118 | 9,998 ± 891 | 61.1 | 83.9 |

GET throughput is approximately 10,000 MiB/s (~84 Gbps) across all object sizes, consistent with the single-node VIP delivering near the effective bandwidth of one application node's bond. PUT throughput ranges from 5,576 MiB/s at 4 MiB to 7,369 MiB/s at 512 MiB. The lower PUT throughput at small object sizes reflects per-request overhead: MinIO's erasure coding, metadata writes, and acknowledgement round-trips add a fixed cost per object that dominates at 4 MiB but amortises at larger sizes.

GET throughput is notably insensitive to object size — the 4 MiB and 2 GiB results are within 5% of each other. This is significant for EO workloads that mix Zarr chunk access (4 MiB) with bulk transfers (512 MiB, 2 GiB): the fabric delivers consistent read bandwidth regardless of access granularity.

### 6.3 Distributed Multi-Client Performance

Table 9 reports distributed throughput with three clients and 96 concurrent connections in total.

Table 9. Distributed multi-client MinIO throughput (3 warp clients on the application tier, 32 concurrent connections each for 96 in total, addressing all 8 storage nodes directly).

| Size | PUT (MiB/s) | GET (MiB/s) | PUT (Gbps) | GET (Gbps) |
|---|---|---|---|---|
| 4 MiB | 13,218 ± 57 | 18,217 ± 164 | 110.9 | 152.8 |
| 64 MiB | 16,092 ± 168 | 18,867 ± 520 | 135.0 | 158.3 |
| 512 MiB | 15,954 ± 268 | 17,946 ± 531 | 133.8 | 150.5 |
| 2 GiB | 16,544 ± 343 | 17,287 ± 1,800 | 138.8 | 145.0 |

PUT scales effectively from single to distributed: 2.37× at 4 MiB, 2.16× at 512 MiB. At larger object sizes, distributed PUT reaches 16,544 MiB/s (139 Gbps) at 2 GiB. GET stays in the range 17,287–18,867 MiB/s across the four object sizes (Table 9).

**st06 configuration anomaly.** During initial benchmarking, a consistent performance anomaly was identified on storage node st06 during distributed GET operations. While PUT throughput to st06 was balanced with the rest of the cluster (within 0.1%), GET throughput from st06 was severely degraded — as low as 137 MiB/s at 4 MiB compared to 2,500–3,200 MiB/s for other nodes. Diagnostic investigation identified the root cause as TCP Selective Acknowledgement (SACK) being disabled on st06 (net.ipv4.tcp_sack = 0) while enabled on all other nodes. Without SACK, any packet loss during reads forces go-back-N retransmission rather than selective recovery. Following correction of the configuration, the distributed GET experiment was re-run. The results in Table 9 reflect the post-correction data; st06 now performs on par with the top-performing nodes in the cluster (2,214–2,414 MiB/s depending on object size).

The per-host GET throughput distribution in the post-correction data shows no systematic node-pair grouping; per-size spreads range from 7% to 20% across the four object sizes — larger than the PUT spreads reported above, but consistent across hosts. Earlier benchmarks of the same workload had shown a more pronounced asymmetry; that asymmetry was traced to boot-time configuration drift in the TCP congestion-control algorithm across four of the eight storage hosts, remediated prior to the data collection reported in Table 9. The diagnostic narrative is recorded in §8.4.

These findings illustrate a broader point: the benchmarking process itself surfaced configuration inconsistencies that were invisible during normal operations. The systematic, per-host analysis made them detectable; without it, the affected nodes would have silently reduced aggregate GET throughput without any visible error.

### 6.4 Mixed Read/Write Performance

Table 10 reports throughput under simultaneous GET and PUT workloads.

Table 10. Mixed read/write throughput (50/50 request split, 3 warp clients × 32 concurrent connections = 96 in total).

| Size | GET (MiB/s) | PUT (MiB/s) | Combined (MiB/s) | GET:PUT ratio |
|---|---|---|---|---|
| 4 MiB | 13,838 ± 373 | 4,614 ± 126 | 18,453 | 3.0:1 |
| 512 MiB | 14,300 ± 358 | 4,785 ± 126 | 19,085 | 3.0:1 |

Combined throughput under mixed workloads (18,453–19,085 MiB/s) is comparable to the distributed GET-only or PUT-only numbers, indicating that MinIO handles concurrent read/write operations without significant mutual interference. The 3:1 GET:PUT throughput ratio under a 1:1 request ratio reflects the asymmetry between read and write paths: GET operations complete in a single round-trip, while PUT operations require erasure coding computation and acknowledgement from multiple storage nodes before returning.

For an operational EO facility — where satellite data ingestion (PUT) runs continuously alongside user queries and downloads (GET) — this result confirms that concurrent access does not degrade either operation below usable levels. The aggregate bandwidth is shared, not serialised.

### 6.5 Network-Throttled Baselines

Table 11 is the central result of this paper: object storage throughput at three network speeds on identical hardware.

Table 11. MinIO cluster throughput on three network configurations (identical hardware). The "100 Gbps" column is the unthrottled distributed result on the deployed 100 GbE fabric; the "10 Gbps" and "1 Gbps" columns are the throttled distributed result simulating each of the three application servers with a 10 Gbps or 1 Gbps NIC respectively, applied as bidirectional tc shaping. The 1 Gbps configuration corresponds to the commercial networking proposal rejected during the AIR Data Centre's procurement.

| Size | Operation | 100 Gbps (MiB/s) | 10 Gbps (MiB/s) | 1 Gbps (MiB/s) | 10G/100G | 1G/100G |
|---|---|---|---|---|---|---|
| 4 MiB | PUT | 13,218 | 2,331 | 233 | 17.3% | 1.7% |
| 4 MiB | GET | 18,217 | 2,291 | 219 | 12.1% | 1.2% |
| 64 MiB | PUT | 16,092 | 2,330 | 233 | 12.7% | 1.3% |
| 64 MiB | GET | 18,867 | 2,306 | 229 | 12.3% | 1.2% |
| 512 MiB | PUT | 15,954 | 2,334 | 232 | 12.4% | 1.2% |
| 512 MiB | GET | 17,946 | 2,347 | 226 | 13.1% | 1.3% |

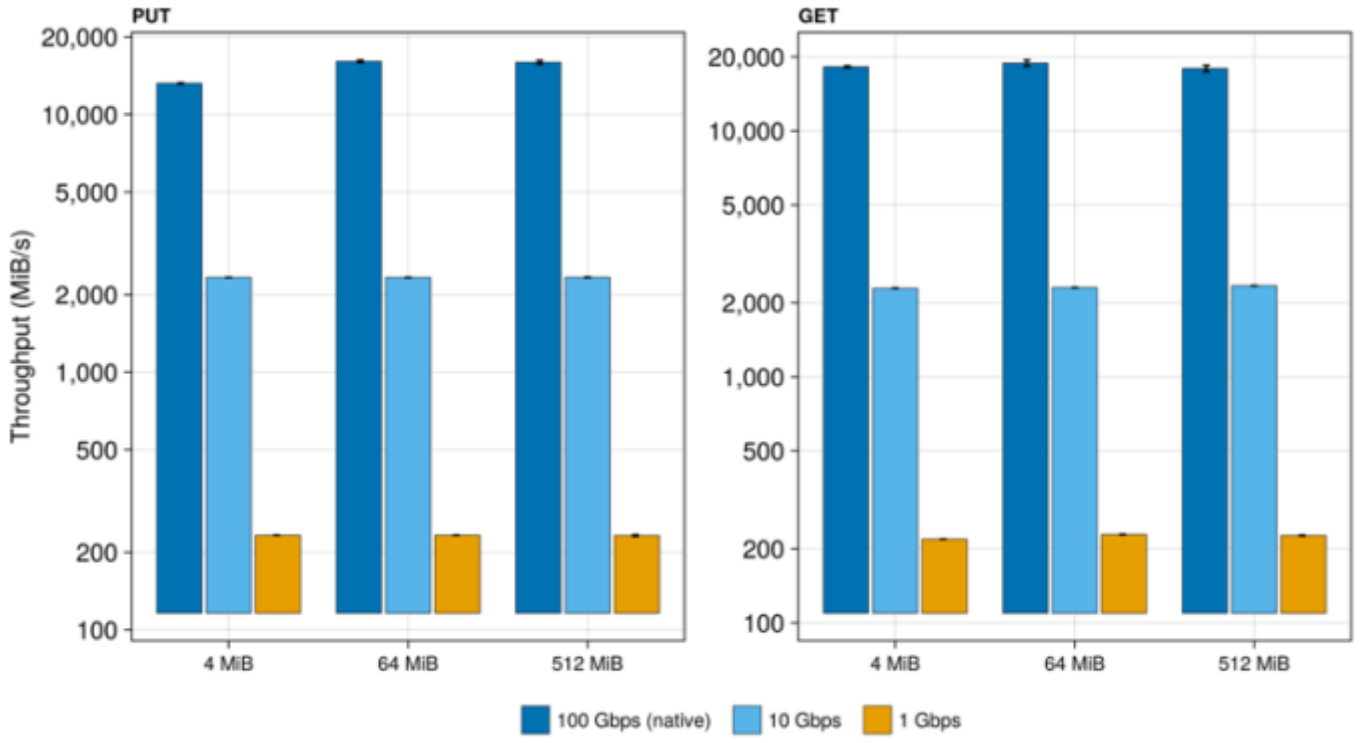

Fig. 7. MinIO cluster throughput on the deployed 100 GbE fabric and on two simulated alternative architectures (each application server with a 10 Gbps or 1 Gbps NIC, modelled by per-server tc shaping), for PUT and GET operations across object sizes. Y-axis is logarithmic; whiskers show ±1 standard deviation. See Table 11.

The throttled configurations simulate the alternative architecture that was proposed for the AIR Data Centre during procurement: each application server fitted with a single 1 Gbps NIC. Under that configuration, the cluster would have sustained approximately 225 MiB/s of aggregate payload throughput (~1.9 Gbps) — what users of a "traditional" 1 Gbps-per-server EO data facility would experience. With 10 Gbps NICs in each server, aggregate throughput would have reached approximately 2,310 MiB/s (~19 Gbps). On the deployed 100 GbE fabric, the cluster sustains approximately 17,000 MiB/s (~140 Gbps). The three configurations span two orders of magnitude in user-perceived performance on identical hardware.

The practical implications follow directly from these numbers. A single 512 MiB Sentinel-2 granule would have taken approximately 2.3 seconds to transfer at 1 Gbps per server, and approximately 218 milliseconds at 10 Gbps per server; on the 100 GbE fabric, it completes in approximately 29 milliseconds. At 1 Gbps per server, interactive browsing of EO archives would have been impractical — each granule access incurring a multi-second wait, and the iterative spatial-subsetting workflows that motivate cloud-native EO formats (Zarr, COG) collapsing into a paged interface to slow storage. On the deployed 100 GbE fabric, the same operation is perceptually instantaneous, enabling the iterative, exploratory workflows those formats were designed for.

The throttled rows of Table 11 are not a network-degradation experiment but a direct measurement of the alternative the AIR Centre rejected.

The throttled configurations also confirm that the cluster is network-bound below 10 Gbps per server. Measured throughput in both throttled configurations is approximately 63% of the cluster's aggregate ceiling at the throttled rate (358 MiB/s at 1 Gbps per server, 3,576 MiB/s at 10 Gbps per server); the consistent ratio across two orders of magnitude of bandwidth, together with the negligible effect of object size at throttled rates (Fig. 7), is the signature of a network-bound regime. Above 10 Gbps per server, throughput no longer scales linearly with the per-server rate — the 10G/100G ratios in Table 11 are 12–18% rather than the 10% that pure linear scaling would predict. Because the throttled rates tested are 1 and 10 Gbps per server against a deployed 100 Gbps, the transition is bounded between 10 and 100 Gbps rather than located within that interval; intermediate rates would narrow it.

### 6.6 Time-to-First-Byte

TTFB at 100 Gbps is 2.5–6.3 ms, scaling with object size — consistent with MinIO's read path initiating data transfer before the entire object is read from disk. Under throttled conditions, TTFB increases to 8–12 ms, reflecting TCP congestion control's slower ramp-up at reduced bandwidth. The TTFB differences between 10 Gbps and 1 Gbps are small because TTFB measures time to the first byte, not throughput — the initial response is fast regardless of the sustained transfer rate that follows (Table 12).

For Zarr-based EO workflows where a single analysis may issue thousands of small chunk requests, TTFB is the dominant performance metric. The sub-3 ms TTFB at 100 Gbps means that 1,000 sequential Zarr chunk reads would accumulate less than 3 seconds of latency overhead, making interactive exploration of multi-dimensional EO datasets practical.

Table 12 reports time-to-first-byte (TTFB) for GET operations across configurations.

Table 12. Mean time-to-first-byte for GET operations (ms).

| Size | Single (100G) | Distributed (100G) | Throttled (10G) | Throttled (1G) |
|---|---|---|---|---|
| 4 MiB | 2.9 | 2.5 | 11.6 | 8.3 |
| 64 MiB | 3.3 | 3.1 | 10.5 | 9.7 |
| 512 MiB | 4.3 | 4.0 | 11.0 | 9.7 |

| 2 GiB | 6.3 | 5.6 | — | — |
|---|---|---|---|---|

### 6.7 Summary

Object storage throughput scales linearly with network bandwidth until storage I/O becomes the constraint. At 1 Gbps, the network saturates regardless of object size; at 100 Gbps, the cluster delivers 10,000 MiB/s (single client) to 19,000 MiB/s (distributed) before endpoint memory bandwidth limits further scaling. Mixed read/write workloads sustain aggregate throughput comparable to single-operation benchmarks, confirming that concurrent ingestion and retrieval do not degrade performance. The throttled baselines provide the empirical reference for institutions operating at 1 or 10 Gbps: multiply the 1 Gbps numbers by your available bandwidth to estimate achievable throughput, with the caveat that returns diminish above 10 Gbps for this hardware class. These results, combined with the network characterisation in §5, confirm the paper's central claim: for on-premises EO data infrastructure, the network investment determines the performance ceiling.

## 7. Remote Access from Partner Institutions

The measurements reported in §5–§6 characterise the AIR Data Centre as a single node: its network fabric, object storage cluster, and metadata catalogue under local load. The Atlantic Cloud model is predicated on access from geographically distributed sites, where the binding constraint is expected to be the public internet rather than the local fabric. To quantify end-user experience beyond the host site, a remote-benchmarking campaign was conducted across partner institutions in the Atlantic basin, with a sample of public-cloud regions included as a comparator. §7.1 documents the methodology; §7.2 reports a first round of results across 63 sites.

### 7.1 Methodology

The remote-benchmarking campaign follows a participant-run protocol: each partner site clones a shared scripted workflow, provides site-specific configuration (credentials, site identifier), and submits results through a version-controlled repository. This lowers the operational burden on partners and ensures that every site executes the same measurement procedure.

The campaign characterises MinIO multi-site replication between a partner-side MinIO instance and the AIR Data Centre, using a containerised local server that auto-configures the replication relationship.

The campaign measures one quantity. Replication throughput is characterised by uploading objects at matching sizes to a source bucket at the partner site and timing the replication window until the object is visible at the AIR Data Centre. Integrity is verified by comparing MD5 checksums of source and replicated objects. Each replication run polls the destination for a visible object at two-second intervals up to a 600-second ceiling; replications that do not complete within that window are recorded as censored, with a per-rep flag distinguishing successful runs from the censored ones. Throughput for a censored run is bounded above by the size-divided-by-ceiling rate of 0.85 MiB/s for the 512 MiB object size; cells in which all three repetitions are censored are reported as upper bounds (< 0.85 MiB/s annotated (3/3)), and cells with mixed outcomes are reported as the mean of the successful repetitions annotated with the censored count, e.g. (1/3).

Measurements are submitted either as a pull request against a results/<site> branch or as an attachment to a GitHub issue, with each partner site identified by an institution-scoped directory. A sample of public-cloud regions, executed against the same measurement scripts, is included as a comparator to contextualise partner-site results against a widely referenced baseline.

### 7.2 Results

A first round of multi-site replication benchmarks has been collected across 63 sites and is reported here as the empirical setup for §8.5. Each (site, size) cell summarises three repetitions; the censoring convention introduced in §7.1 carries through this section. The institutional set covers seven sites — NCICS (NOAA, USA), IFF (Brazil), LAMCE (UFRJ, Brazil), NASRDA (Nigeria), LANOT (UNAM, Mexico), the University of Waterloo (Canada), and TACC (USA) — alongside 17 AWS regions and 39 GCP regions.

#### *7.2.1 Cross-site replication throughput*

*Cross-site replication throughput characterises bucket-to-bucket replication: a partner-side MinIO writes an object, MinIO's replication state machine propagates it to the AIR Data Centre, and the run completes when the destination reports the object visible. Across 567 replication runs (63 sites × 3 sizes × 3 reps), 11 were censored, all at 512 MiB. Three sites — aws/ca-central-1, gcp/australia-southeast1, and NASRDA — are fully censored (3/3) at 512 MiB. Two sites — gcp/asia-south2 and gcp/australia-southeast2 — are partially censored (1/3) at 512 MiB; their measured repetitions report 0.90 and 0.89 MiB/s respectively, immediately above the timeout-implied floor.*

The censored cells fall into two populations, and the distinction governs how they should be read. The cloud regions sit at the low end of the throughput distribution, where replication time at 512 MiB approaches the ceiling and variance pushes the slowest repetitions past it; their censoring is consistent with their own throughput at the smaller object sizes. NASRDA is a different case. It replicates 64 MiB COG tiles at 2.51 MiB/s — 2.67× the median of the cloud regions in its own RTT band, and the second-highest institutional ratio at that size — which projects to roughly 204 s for a 512 MiB object, comfortably inside the 600-second window. All three of its 512 MiB repetitions nevertheless exceeded the ceiling. The censoring is therefore specific to the largest object size and is not explained by the site's sustained throughput at smaller ones. These measurements do not resolve why, and we do not speculate here; the effect is a candidate for the systematic attribution reserved for future work.

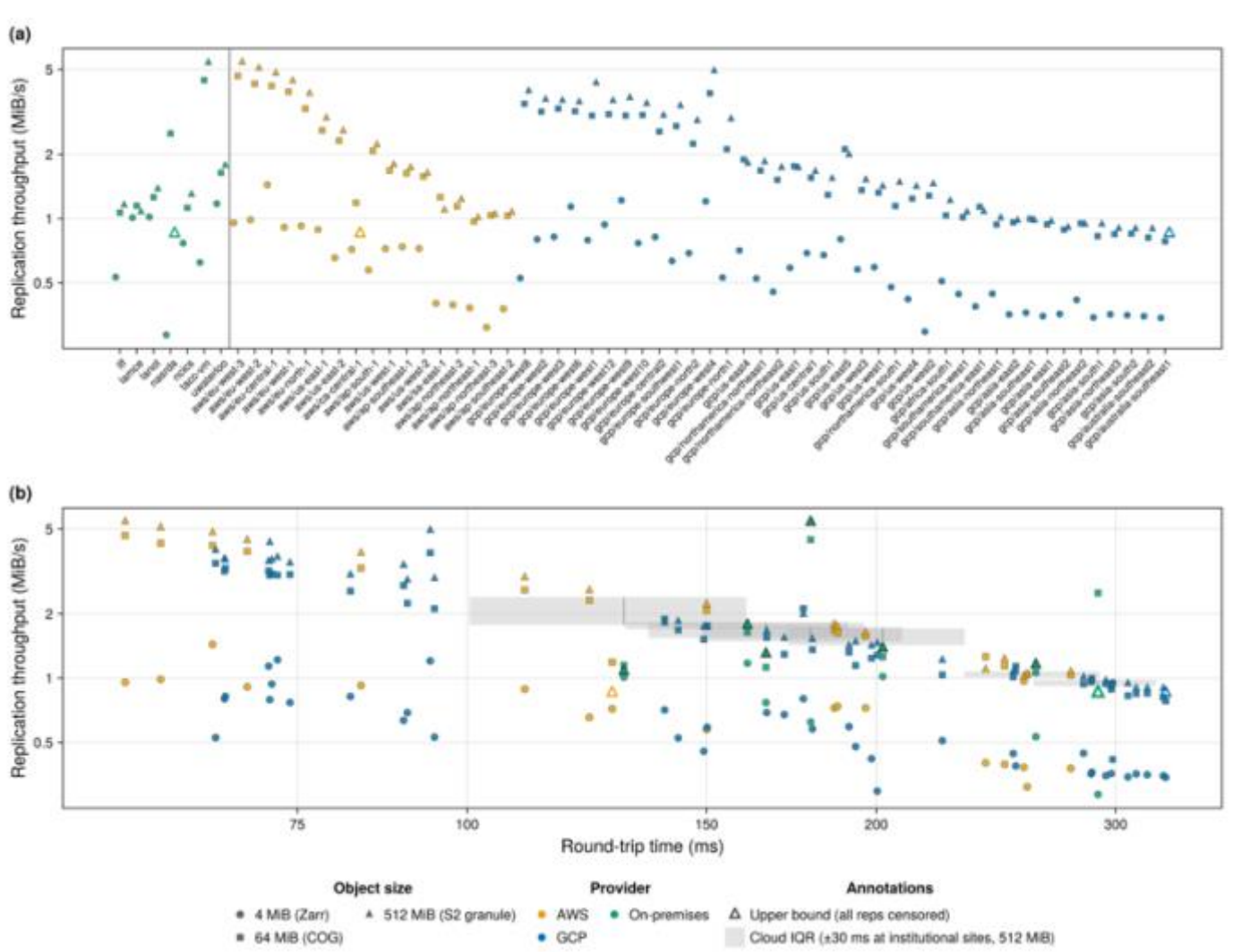

Fig. 8. Multi-site replication throughput. Panel (a): per-site replication throughput at 4 MiB, 64 MiB, and 512 MiB; sites are ordered with institutional partners first

(alphabetical) and cloud regions following, grouped by provider and RTT-ordered within provider. Censored cells (all three repetitions censored) render as upper-bound markers. Panel (b): RTT vs replication-throughput scatter across all 63 sites; marker shape encodes object size, colour encodes provider. Shaded grey bands at each institutional site's RTT show the cloud-region IQR within ±30 ms at 512 MiB, to make the within-RTT-band comparison visible.

The dominant visual story in Fig. 8, Panel (a), is not the RTT correlation but the contrast between the institutional cluster on the left and the cloud-region cluster to its right, a contrast that holds for most but not all of the institutional set. The five AWS Europe regions (56–84 ms RTT) replicate at 3.86 to 5.41 MiB/s on 512 MiB; the GCP Europe block in the same RTT range sits at 2.88 to 4.93 MiB/s. Most institutional sites at substantially higher RTT replicate below this Europe-side band: the University of Waterloo (160.8 ms) at 1.78 MiB/s, LANOT (202.2 ms) at 1.38 MiB/s, NCICS (165.9 ms) at 1.30 MiB/s, IFF (262.0 ms) at 1.16 MiB/s, LAMCE (130.4 ms) at 1.08 MiB/s, and NASRDA (291.2 ms) bounded below 0.85 MiB/s. TACC (179.0 ms) is the exception, replicating at 5.39 MiB/s — inside the AWS Europe range and above every GCP Europe region. The relevant comparator for an Atlantic-basin institutional partner is not AWS Europe but the cloud region at comparable RTT. Within ±30 ms RTT bands, six of the seven institutional sites exceed the cloud-band median on 4 MiB Zarr chunks — LANOT and the University of Waterloo by 1.74×, LAMCE by 1.63×, IFF by 1.38×, NCICS by 1.30×, and TACC by 1.06× — while NASRDA reaches 0.80× of its band median. IFF and TACC exceed their band medians at all three object sizes. At 64 MiB and 512 MiB the other institutional sites fall below their band medians by factors between 0.58 and 0.98: at most 1.7× slower than cloud at the same RTT, and within single-digit minutes per granule at 512 MiB — 1.6 minutes at TACC, 4.8 at the University of Waterloo, 6.2 at LANOT, 6.6 at NCICS, 7.4 at IFF, and 7.9 at LAMCE, against cloud-band medians of 4.6 to 8.9 minutes. NASRDA, censored at 512 MiB, requires at least 10.0 minutes. TACC is the instructive case: at 179 ms it is further from the AIR Data Centre than either LAMCE (130 ms) or the University of Waterloo (161 ms), yet at 512 MiB it replicates at 3.27× its own cloud-band median and faster than every GCP Europe region at a third of the distance. Two features of TACC's row are not explained by network class alone: it is the slowest institutional site at 4 MiB while the fastest at 512 MiB, and the site runs as a virtual machine on TACC infrastructure, which is not characterized here. The site is therefore an illustration of the pattern rather than evidence for it; the 63-site comparison carries the claim. LANOT makes the same point more modestly: at 202 ms it replicates faster than both LAMCE and NCICS (166 ms) and comes within 8% of its cloud comparators at 512 MiB. Distance alone does not set the ceiling.

The institutional/cloud divide is visible in replication at all three object sizes but is sharpest at 512 MiB, where the timeout boundary makes the floor explicit. The class of network the path traverses, rather than RTT, is what sets the ceiling; within the cloud set itself, AWS Europe medians exceed GCP Europe medians in replication by 20–37 % depending on object size (Table 13). The AIR Data Centre as a replication target is operational across the institutional set with research-network connectivity at 512 MiB and bounded by the timeout for the institutional case without it.

#### *7.2.2 Summary*

Table 13 summarises replication throughput at the three EO object sizes for the institutional set and for cloud-region groupings. Institutional rows report the per-site mean of three repetitions; cloud-region rows report the median across sites in the group. Censored cells are flagged per §7.1.

Table 13. Multi-site replication throughput by site and region group at three EO object sizes (4 MiB Zarr chunk, 64 MiB COG tile, 512 MiB Sentinel-2 granule). Cell entries are MiB/s, reported as the mean of successful repetitions of three. Cells with the (3/3) annotation are fully censored: the value is an upper bound, not a measurement, and equals the object size divided by the 600-second poll ceiling. Group medians for the cloud-region rows aggregate the per-site means listed in the underlying CSVs (analysis/path_b_replication_throughput.csv). Asterisked 512 MiB cells (*) report the median across unbound cells only; the AWS — North America row excludes one fully-censored cell (aws/ca-central-1) and the GCP — Asia-Pacific + South America row excludes one fully-censored cell (gcp/australia-southeast1). Two further sites (gcp/asia-south2, gcp/australia-southeast2) are partially censored at 512 MiB; their per-site means in the underlying CSV report the mean of the two successful repetitions and are included in the medians.

| Site / region group | n | 4 MiB | 64 MiB | 512 MiB |
|---|---|---|---|---|
| NCICS (NOAA, USA) | 1 | 0.77 | 1.12 | 1.30 |
| IFF (Brazil) | 1 | 0.53 | 1.07 | 1.16 |
| LAMCE (UFRJ, Brazil) | 1 | 1.01 | 1.15 | 1.08 |
| NASRDA (Nigeria) | 1 | 0.29 | 2.51 | < 0.85 (3/3) |
| LANOT (UNAM, Mexico) | 1 | 1.02 | 1.26 | 1.38 |
| University of Waterloo (Canada) | 1 | 1.18 | 1.64 | 1.78 |
| TACC (USA) | 1 | 0.62 | 4.45 | 5.39 |
| AWS — Europe | 5 | 0.96 | 4.19 | 4.80 |
| AWS — North America | 5 | 0.72 | 1.68 | 2.18* |
| AWS — Asia-Pacific + South America | 7 | 0.39 | 1.14 | 1.10 |
| GCP — Europe | 13 | 0.80 | 3.06 | 3.57 |
| GCP — North America | 12 | 0.58 | 1.44 | 1.60 |
| GCP — Asia-Pacific + South America | 13 | 0.36 | 0.94 | 0.96* |
| GCP — Africa (south1) | 1 | 0.51 | 1.04 | 1.21 |

#### *7.2.3 Closing observations*

Our principal finding is that replication throughput across all sites is bounded by the type of network the replication path traverses rather than by RTT alone, with the institutional/cloud divide visible at every object size and sharpest at 512 MiB. TACC illustrates it most visibly, with the caveat noted in §7.2.1: at 179 ms RTT it replicates a 512 MiB granule at 5.39 MiB/s, faster than every GCP Europe region at a third of the distance and 3.27× the median of the cloud regions in its own RTT band. LANOT makes the same point at smaller magnitude: at 202 ms it replicates faster than institutional sites at half that distance, and comes within 8% of the cloud regions in its own RTT band at 512 MiB.

## 8. Discussion

The results confirm what we suspected when we rejected the 1 Gbps proposal during procurement, but they also surprised us. That the network sets the ceiling was hypothesised, and the measurements bear it out. What we did not anticipate was the degree to which memory topology, rather than memory capacity, determines how much of that ceiling the system actually reaches. Nor did we set out to use benchmarking as a diagnostic tool, though that turned out to be among the more operationally valuable outcomes of the exercise. This section works through these findings and their implications for other institutions considering similar deployments.

### 8.1 The Network as a First-Class Resource

Most of the storage literature treats the network as invisible. Benchmarks report throughput that depends on an uncharacterised fabric, and the reader is left to assume the network was not the bottleneck. The throttled baselines in §6.5 bring the dependency to the fore: on identical hardware, MinIO PUT throughput at 512 MiB falls from 15,954 MiB/s to 2,334 MiB/s to 232 MiB/s as the per-node link rate drops from 100 Gbps to 10 Gbps to 1 Gbps. A factor of 69 across two orders of magnitude of bandwidth, with nothing else changed.

These numbers have a direct operational reading. At 1 Gbps, pulling a single 512 MiB Sentinel-2 granule takes about 2.3 seconds. That does not sound catastrophic until one considers what EO work actually involves. Researchers do not download one granule; they browse dozens, flipping through scenes to build intuition about a region before selecting the data they need. At 2.3 seconds per granule, that visual workflow is a slideshow. At 29 milliseconds — the measured time on the 100 GbE fabric — it is fluid, closer to scrolling through images on a local disk. The difference is not incremental; it changes which workflows are practical at all. Cloud-native formats such as Zarr and COG were designed around rapid, iterative partial reads. On a 1 Gbps fabric, those formats cannot do what they were designed to do.

### 8.2 Hardware Selection: Spending Where Reversal Is Hardest

Two questions govern any hardware decision taken at the commissioning of a new facility, and neither is how much a component costs. The first is whether the decision can be revisited later: adding memory to a running application node is a routine intervention of a few minutes with the chassis open, using commodity parts bought when they are needed, whereas re-fabricating the network — new switches, new optics, re-cabling between racks, reconfiguring bonding, all inside a live facility — is a multi-year undertaking. On operational timescales the fabric is effectively irreversible and the endpoints are not. The second is what load the purchase must serve at the moment it is made: a new data centre for a new institution has, by definition, almost no load, so provisioning the endpoints for a workload that does not yet exist buys capacity that sits idle and depreciating.

Both questions point the same way, and the AIR Data Centre was built on the answer: spend on the axis that cannot be revisited, and defer on the axis that can. The 100 GbE fabric was bought once, in full, for the ceiling the facility would need across its service life, that being the last point at which the ceiling can be chosen cheaply; the application tier was specified at 32 GB per node across two of sixteen memory channels. The deferral is not a concession the budget forced but the mechanism by which the fabric was afforded, and it cost nothing, because the capacity deferred was capacity nothing yet required. The inverse would have lost on both counts: memory bought early for an absent workload, replaced anyway once real demand arrived, behind a fabric the facility could not economically re-lay.

The benchmarks quantify what the deferred purchase will buy, and this is their value to a procurement decision rather than to a performance claim. The application nodes reach 44% of their 400 Gbps bond under parallel load, bounded by the two-channel memory configuration (§5.2, §5.5); the R750 storage nodes, on the same switches and cabling, reach 86%; and the R7625 virtualisation host, with 24 populated DDR5-4800 channels and 921.6 GB/s of memory bandwidth, sustains 94.5% of its bond in fan-out to the storage tier (§5.5, Table 5b). The fabric is not the ceiling in any of these cases; the endpoint is, momentarily. What separates 44% from 94.5% is memory provisioning, and it is purchasable without touching a switch. The R7625 is the same strategy applied in the other direction: when a workload appeared that justified a large, well-balanced memory configuration, one was bought, and it saturates the fabric the application tier does not yet reach. The 44% is therefore a price tag rather than a fault, measuring what the deferred upgrade will recover on hardware already racked and cabled, at a moment the institution chooses.

For the application tier that upgrade is an in-chassis expansion. Each node has sixteen memory channels and populates two; filling all sixteen raises theoretical memory bandwidth from roughly 51 GB/s to roughly 410 GB/s, an eightfold increase. Combined with an NPS=1 setting — a firmware change carrying no hardware cost — which collapses each socket to a single NUMA domain with all cores local, this brings the application tier to the same memory configuration as the R750 storage nodes that saturate 86% of the bond on the identical fabric. One in-chassis upgrade and one BIOS setting lift the tier to the network's level. Because channel bandwidth is set by channel count rather than module size, the same 410 GB/s is reached with any DIMM size, so module size sets capacity rather than bandwidth. For a facility running memory-intensive scientific workloads, capacity per core is a first-order constraint alongside bandwidth, so the sensible target is modules that also raise capacity: sixteen 16 GB DIMMs deliver 256 GB, or 8 GB per core, the same balance as the R7625, at full channel bandwidth. The upgrade is therefore planned as a single capacity-and-bandwidth step that reaches the configuration the facility will keep, with nothing discarded along the way.

The design returned its value immediately, delivering from deployment the interactive, iterative access to EO data that cloud-native formats assume, and the same purchase also carries forward. The requirement the fabric was sized against is set by how researchers work and which data they base their work on rather than by how large the archive becomes: a granule must arrive in tens of milliseconds whether the archive holds tens of terabytes or several petabytes. Growth in EO data volume therefore lands on storage capacity, not on the per-request responsiveness that determined the network specification. Capacity is added by racking further storage nodes, and the fabric carries considerably more aggregate throughput than the endpoints connected to it draw: the spine sustains 1,379 Gbps under symmetric storage load while the application tier reaches 44% of its own bond. The scale-out property of the storage tier matters here as much as the spare bandwidth, because each added node is a commodity server contributing capacity, memory, cores and four further 100 GbE ports in the same chassis, so an expanded cluster serves the enlarged archive at the resource ratios the original was built to rather than spreading a fixed pool more thinly.

The facility has already absorbed expansion of this kind without touching the fabric. Specified in 2021 against the eight storage and three application nodes then planned, it has since absorbed three further tiers — virtualisation, GPU, and modelling — spanning a processor generation beyond the original nodes, each connected to the switches already in place. No switch was added, no cabling re-laid, no bonding scheme revised. The best-provisioned of those later arrivals sustains 94.5% of a single node's 400 Gbps bond (§5.5, Table 5b), the closest any endpoint in the facility has come to its share of the fabric, and 58 of the 128 spine ports are in use, leaving capacity for roughly seventeen further servers at four ports each (Appendix A). The compute tiers are expected to turn over several times within the fabric's service life, and each turnover becomes a purchase decision taken on its own merits rather than an occasion to reconsider the interconnect. The argument holds while per-node demand remains within what four bonded 100 GbE ports carry and while concurrent load stays within the aggregate the spine sustains.

One qualification is institutional rather than architectural. The modular strategy — build the irreversible fabric in full, grow the reversible endpoints on demand — was rooted on being able to purchase incrementally, timing each upgrade to the load that justified it. Partway through the deployment the procurement regime changed: specifications now have to be fixed annually and purchased in bulk, which constrains the timing flexibility the modular approach was designed around. The sequencing described here remains correct — it would be correct at any budget, and the design is what made the facility affordable at all — but the responsiveness that made deferral efficient is now constrained by how the institution is required to buy, not by anything in the architecture. Institutions replicating this model should weigh their own procurement agility as carefully as their hardware, because the two are not independent.

**8.3 LACP Bonding Under Parallel EO Workloads**

The stream-balance results in §5.4 quantify a behaviour that bonded Ethernet is known to exhibit under high concurrency; what matters for EO workloads is the magnitude. Under no contention, the 8 TCP streams in each iperf3 session distribute reasonably well across the 4 bonded links: 9.6% coefficient of variation. Under full-mesh contention with 24 concurrent sessions, balance collapses to 93.3% CV, with some streams receiving less than 1 Gbps on a 400 Gbps bond.

The mechanism is LACP's hash-based flow assignment. Flows are mapped to physical links by a hash of the source and destination address and port tuple. With many flows sharing few links, hash collisions concentrate traffic on whichever links happen to attract the most flows, and the busiest link becomes the bottleneck regardless of what the other three are doing.

For EO access patterns, this means that a user retrieving Zarr chunks while a colleague ingests Sentinel-2 data will see per-request throughput that varies with hash luck, not only with how busy the cluster is. We have no clean fix for this within the LACP model; it is inherent to hash-based distribution. The pragmatic response is to understand the limitation and account for it in capacity planning: effective per-client bandwidth under parallel load is less than the aggregate divided by the number of clients. The storage-to-storage mode, where both endpoints have higher memory bandwidth, shows 22.8% stream CV — better than the 93.3% under full mesh, because fewer flows compete per link when each flow can individually saturate more of the bond. This is another respect in which the endpoint upgrade described in §8.2 would improve not only throughput but its predictability.

RDMA over Converged Ethernet (RoCE) offers a different flow model that would reduce the effect, but at higher cost and operational complexity than most institutional deployments can justify. For the majority of institutional EO deployments, LACP bonding on commodity Ethernet remains the right trade-off, provided its behaviour under contention is understood rather than assumed away.

**8.4 Benchmarking as a Diagnostic Tool**

We did not plan for benchmarking to be a diagnostic exercise, but it became one. Several issues that were invisible under normal operations surfaced only under the controlled, per-node measurement protocol:

A BIOS configuration error on storage node st04 (CPU frequency boost disabled) reduced its throughput to 84 Gbps — 22% below the cluster mean. The deficiency was undetectable during normal MinIO operations because the load balancer distributed requests across all nodes; the degraded node simply contributed less without triggering any error. Per-pair throughput benchmarking exposed the anomaly immediately.

Docker Swarm workloads on application node srv01 — including the Traefik reverse proxy and single-replica services — caused memory contention under parallel load, producing 1,700 retransmits while the other application nodes had near-zero. Draining Swarm services before benchmarking eliminated the issue, confirming that production workloads on benchmarking nodes introduce confounds that can mask or mimic hardware problems.

TCP SACK (Selective Acknowledgement) was disabled on storage node st06, reducing GET throughput by 95% for small objects. The configuration inconsistency affected only the read path and only under parallel load — conditions that normal operations rarely stress uniformly across all nodes. Per-host analysis of distributed GET benchmarks made the anomaly visible.

TCP congestion-control configuration on four of the eight storage nodes (st01, st02, st07, st08) had drifted from BBR to cubic, while the remaining four (st03–st06) ran BBR as intended. The split produced a two-tier per-host GET throughput pattern during distributed benchmarks that ran for hundreds of seconds, where long-lived MinIO peer-to-peer connections settled into asymmetric congestion behaviour. Production hid the inconsistency because both algorithms function correctly in isolation and the load balancer averaged across hosts; only systematic per-host analysis under sustained parallel load surfaced the asymmetry. Per-connection congestion-control inspection during a live distributed PUT confirmed the divergence, and live `tc qdisc` adjustments had no effect on per-host throughput — ruling out queueing discipline and pointing to the kernel-level TCP state.

None of these problems would have surfaced without systematic, per-node benchmarking under controlled conditions. For other institutions, the benchmark suite is not only a publication artefact; it is an operational baseline. Run it after deployment, record the results, and re-run it after any kernel update, firmware change, or hardware intervention. The cost is a few hours of maintenance-window time. The return is catching configuration drift before it silently degrades the system.

**8.5 Relevance to Distributed EO Infrastructure**

The Atlantic Cloud model described in §3 depends on each node being independently capable of serving EO workloads at useful performance levels. The benchmarks in §5–§6 establish what "useful" means at the host site: a single client retrieving data through the API can expect approximately 10,000 MiB/s on the 100 GbE fabric, and the cluster sustains approximately 225 MiB/s aggregate when each application node is throttled to 1 Gbps — about 75 MiB/s per application node, against an aggregate ceiling of roughly 358 MiB/s. For an institution in West Africa or the Caribbean considering its first on-premises EO data facility, these numbers establish the deployment baseline against which the partner's available network determines what their users would experience.

The remote benchmarks in §7 establish the second part of the picture: what the AIR Data Centre delivers to geographically distributed partners through multi-site replication, and what it implies for federation. The §5–§6 ceiling of approximately 84 Gbps to a single client is the throughput a local Atlantic Cloud deployment delivers to its users today; the §7 results characterise what passes between Atlantic Cloud nodes.

MinIO multi-site replication is operational across the institutional set at 4 MiB and 64 MiB, and at 512 MiB at every institutional site except NASRDA, whose 512 MiB cell is censored. Institutional throughput is comparable to public-cloud egress at the same RTT. Within ±30 ms RTT bands, six of the seven sites exceed the cloud-region median on 4 MiB Zarr chunks, and IFF and TACC exceed it at all three object sizes. The others fall below the cloud-region median on the larger sizes by factors between 0.58 and 0.98 — at most 1.7× slower than the cloud comparator at the same RTT, and within single-digit minutes per 512 MiB granule (1.6 to 7.9 minutes, against a cloud-band median of 4.6 to 8.9 minutes at comparable RTT). Replication is asynchronous; at these rates a single institutional partner can replicate from tens to a few hundred 512 MiB Sentinel-2 granules into the AIR Data Centre per overnight window. For the Atlantic Cloud, federation through replication works across transatlantic links, at rates within a small constant factor of public-cloud egress at the same RTT.

The federation primitive is the architectural mechanism on which the Atlantic Cloud's premise rests: cross-institutional data sharing without dependency on commercial cloud egress, off-site backup between institutions that retain control of their own data, and sovereignty over EO archives that institutions are increasingly required to maintain. That federation works at rates comparable to public-cloud egress, while preserving the sovereignty, governance, and cost structure of an inter-institutional architecture, is the result that supports the Atlantic Cloud as a viable alternative to routing institutional EO data through commercial cloud infrastructure. As future Atlantic Cloud nodes deploy onto research-network connectivity, the path between two Atlantic Cloud nodes shifts from "two ends on the public internet" to "two ends on the same research-network class," and the federation case strengthens correspondingly.

The descriptive cut here cannot resolve where on the partner side the replication throughput ceiling is set: the §7 measurements characterise only the AIR-DC end of each transfer, not the partner endpoint. Candidate factors include partner-side hardware (object-store implementation, NIC capacity, and storage-tier IOPS at the partner endpoint, none of which are characterised in §7), the replication state machine's internal queueing and retry behaviour, and the discrete polling interval used by the measurement script. A systematic attribution is reserved for a companion paper that exploits the same dataset in greater depth.

The 63-site coverage in §7 — seven institutional sites, plus a 56-site sample of public-cloud regions used as a comparator — establishes the empirical setup for federation across the Atlantic basin. As additional Atlantic Cloud nodes come online, the same measurement framework can be extended to characterise inter-node replication directly, beyond the institutional-to-AIR-DC topology measured here.

### 8.6 Limitations

The application tier has three nodes, which caps client-side parallelism; more client nodes would produce different contention and different fabric utilisation. All measurements come from one deployment, so the numbers reflect the hardware, firmware and software versions at the AIR Data Centre. The same architecture elsewhere should behave qualitatively alike — network-bound at high bandwidth, storage-bound at low — with different absolute throughput. The benchmarks also measure raw storage throughput rather than application-level performance: real workflows add API routing, authentication, format parsing and application logic, so these figures are an infrastructure ceiling rather than an end-user experience, and the size of that gap is not quantified here.

The throttled baselines vary two things, not one: bandwidth, and client concurrency, which was reduced from 32 connections to 4 at 1 Gbps to avoid MinIO multipart timeouts on 512 MiB objects. Bandwidth dominates — throughput at 1 Gbps is within a few per cent of one-tenth of the 10 Gbps figure at every object size and for both operations, which is what a bandwidth-bound system predicts, and the lower concurrency would if anything depress it further. The two cannot be fully separated from these measurements, but separation was not the objective: the throttled configuration reproduces the 1 Gbps core rejected at procurement, reduced concurrency included.

Finally, we report no cost data. Procurement costs are contract-specific and would date quickly, so what this paper supplies is the performance side of a cost-performance analysis, not the analysis itself.

## 9. Conclusion and Future Work

This paper presented the design and empirical evaluation of a network-centric, on-premises architecture for Earth observation data access, built on S3-compatible object storage (MinIO), spatial metadata management (PostGIS) and standard OGC API access, and evaluated at its first operational deployment in the Azores. The 100 GbE fabric sustains 86% of theoretical capacity under symmetric load, so the fabric itself is not the binding constraint. On it, distributed GET throughput holds across object sizes from 4 MiB to 2 GiB, and retrieving a 512 MiB Sentinel-2 granule spans a 79× range between 1 Gbps and 100 GbE on identical hardware.

The implication is practical. For institutions planning on-premises EO infrastructure, these results supply the empirical reference the literature has lacked: measured throughput at three network speeds, for EO-representative object sizes, on commodity hardware. The network — not storage capacity, not compute — is the decision with the largest impact on what the system can deliver, and the longest reach. The fabric specified in 2021 delivered interactive access from the day the facility opened and has since absorbed three additional server tiers without modification, retaining both spare ports and unsaturated per-node capacity. An institution sizing a first facility is not choosing a network for the servers it is buying now, but for the several generations that will follow them.

For the Atlantic Cloud, this paper establishes the performance baseline for the reference node architecture and lowers the barrier for the next institution that needs on-premises EO capacity. Partner institutions across the basin have completed a first round of multi-site replication benchmarks, with further sites under discussion. The architecture is documented, the benchmark methodology published, and the network open to new nodes.

Several directions follow. Inter-node replication is the natural extension as the network grows. These benchmarks used synthetic workloads; characterising performance under production traffic — ingestion from the Direct Receiving Station, OGC API queries, cross-site replication — would test them against operational reality. Collaboration with TACC on federated governance for software provenance would add an institutional trust layer. A PostGIS and MongoDB comparison at archive scale, and direct external-client access from partner sites and public-cloud regions against an on-premises baseline, are reported separately.

**Appendix A. Facility hardware and network detail**

This appendix collects facility detail that supports reproducibility without bearing on the benchmark results: the two server tiers not exercised by the evaluation, and the network provisioning and staff connectivity of the AIR Data Centre. The full five-tier inventory is given in Table 1; the storage, application, and virtualisation tiers exercised by the benchmarks in §5–7 are described in §4.2.

**GPU tier (1 node).** One Dell PowerEdge R760xa with dual Intel Xeon Gold 5420+ processors (28 cores / 28 threads per socket, 2.0 GHz), 512 GB DDR5-4400, and four NVIDIA L40 GPUs (48 GB VRAM each). This server supports machine learning workloads — both operational inference services, model training, and exploratory research via partner VMs. Per-pair throughput benchmarks of the R760xa are in progress; GPU-tier throughput characterises fabric performance for AI and ML workloads.

**Modelling tier (1 node).** One Dell PowerEdge R7615 with a single AMD EPYC 9684X processor (96 cores / 96 threads, 2.55 GHz, 1152 MiB L3 V-cache) and 256 GB DDR5-4800 RAM, with four 1.92 TB NVMe drives. This server is dedicated to computationally intensive modelling workloads — CFD, ocean circulation models — and is available to Atlantic Cloud partners. Per-pair throughput benchmarks of the R7615 are in progress and will be reported alongside the application- and storage-tier results.

**Network provisioning.** The storage and application tiers use direct-attach copper (DAC) cables; the newer servers (virtualisation, GPU, modelling) use fibre, which became more affordable and is easier to manage in constrained rack space. The 14-server deployment uses 56 of the 128 available 100 GbE ports; a further 2 ports provide uplinks to the 10 GbE aggregation switch described below, bringing the total to 58 ports in use and leaving 70 ports available for future expansion — sufficient for approximately 17 additional servers at 4 ports each.

**Staff network.** A Dell S4128F-ON 10 GbE switch connects to the 100 GbE fabric via two 100 Gbps uplinks, providing aggregated access for staff workstations. Twenty-four fibre links run directly from this switch to individual desks, each delivering 10 Gbps via QNAP adapters; 12 are currently connected. This arrangement replaced an earlier configuration in which desk connections consumed 100 GbE switch ports directly, freeing those ports for new server deployments.

---

**CRediT authorship contribution statement**

João Pinelo: Conceptualization, Methodology, Software, Formal analysis, Data Curation, Visualization, Writing - Original Draft, Writing - Review & Editing, Project administration, Supervision, Funding acquisition. João Gonçalves: Investigation, Resources, Software, Validation, Writing - Review & Editing. Denis Willett: Software, Investigation, Resources, Writing - Review & Editing. Amit Ruhela: Investigation, Resources, Writing - Review & Editing. Derek Steinmoeller: Investigation, Resources, Writing - Review & Editing. Uriel Mendoza: Investigation, Resources, Writing - Review & Editing. Pelumi S. Alao: Investigation, Resources. Ronald Soares Lopes: Investigation, Resources, Writing - Review & Editing. Rogerio Atem de Carvalho: Resources, Writing - Review & Editing. Pedro Mattos: Investigation, Resources.

**Declaration of competing interest**

The authors declare that they have no known competing financial interests or personal relationships that could have appeared to influence the work reported in this paper.

**Data availability**

The benchmark data, analysis code and derived results supporting this study are openly available in Zenodo [16] at https://doi.org/10.5281/zenodo.22032258, and are developed openly at https://github.com/AIRCentre/network-centric-eo-data-access-paper. Every numerical claim in the paper is traceable to its source measurement through analysis/results/paper_values.csv and analysis/PROVENANCE.md.

**Funding**

The AIR Data Centre was funded principally by the AIR Centre from institutional resources, with partial support from the Regional Government of Azores and *Agenda Mobilizadora* New Space Portugal, as part of Portugal's Recovery and Resilience Plan (RRP) - Project nº 02/C05- i01.01/2022.PC644936537-00000046; IAPMEI Project Nº11. The funders had no role in the study design, in the collection, analysis or interpretation of data, in the writing of the report, or in the decision to submit the article for publication.

**Declaration of generative AI and AI-assisted technologies in the writing process**

During the preparation of this work the authors used large language models to assist with drafting and editing prose. All experimental design, scientific decisions, and final content are the authors'.